\documentclass[10pt,twocolumn,twoside]{IEEEtran}

\IEEEoverridecommandlockouts                              

\usepackage{amsfonts}
\usepackage{amssymb}
\usepackage{amsmath,amsthm,mathtools}
\usepackage{graphicx}
\usepackage{hyperref,caption,subcaption,overpic}

\usepackage{amsopn}
\usepackage{enumitem}

\DeclareGraphicsExtensions{.pdf,.jpeg,.jpg,.png,.eps}
\usepackage{hyperref}

\usepackage{ifthen}
\usepackage[space]{cite}
\DeclareMathOperator{\diag}{diag}
\DeclareMathOperator{\rank}{rank}

\newcommand{\R}{{\rm I\!R}}
\def\rep#1{(\ref{#1})}
\def\scr#1{{\cal #1}}
\newtheorem{thm}{Theorem}
\newtheorem{lem}{Lemma}
\newtheorem{prop}{Proposition}

\newtheorem{cor}{Corollary}
\newtheorem{assum}{Assumption}
\newtheorem{rem}{Remark}

\newcommand{\1}{\mathbf{1}}
\newcommand{\0}{\mathbf{0}}

\title{\LARGE \bf 
Multi-Layer SIS Model with an Infrastructure Network
}

\author{
Philip E. Par\'e,
 Axel~Janson, Sebin~Gracy, Ji Liu,
Henrik Sandberg, and 
Karl H. Johansson
\thanks{Philip E. Par\'{e} is with the School of Electrical and Computer Engineering, Purdue University, IN, USA (
philpare@purdue.edu).}

\thanks{
Axel Janson, Sebin Gracy,       Henrik Sandberg and
        Karl H. Johansson are with the 
        Division of Decision and Control Systems, School of Electrical Engineering and Computer Science, KTH Royal Institute of Technology, and Digital Futures, Stockholm, Sweden.
        (
        axejan@kth.se, gracy@kth.se, hsan@kth.se, kallej@kth.se)
}

\thanks{Ji Liu is with the Department of Electrical and Computer Engineering, Stony Brook University, USA (\texttt{ji.liu@stonybrook.edu})}

\thanks{This work was supported in part by the Knut and Alice Wallenberg Foundation, Swedish Research Council under Grants~2016-00861 and~2017-01078, and the National Science Foundation, grants NSF-CNS \#2028738 and NSF-ECCS \#2032258.}

}

\usepackage[dvipsnames]{xcolor}

\newcommand{\seb}[1]{%
{\leavevmode\color{black}#1}%
}

\newcommand{\sebcancel}[1]{%
{\leavevmode\color{black}#1}%
}

\begin{document}

\maketitle

\begin{abstract}
This paper deals with the spread of diseases over both a population network and an infrastructure network. We 
develop a layered networked spread model for a susceptible-infected-susceptible (SIS) pathogen-borne disease spreading over a human contact network and an infrastructure network, and refer to it as a layered networked susceptible-infected-water-susceptible (SIWS) model. The ``W'' in SIWS represents any
infrastructure network contamination, not necessarily restricted to a water distribution network. 
 We say that the  SIWS network is in the \emph{healthy state} (\sebcancel{also referred to as the disease-free equilibrium}) 
if none of the individuals in the population are infected nor is the infrastructure network contaminated; otherwise, we say that the network is in the \emph{endemic state} \sebcancel{(also referred to as the endemic equilibrium)}. 
\sebcancel{First, we establish sufficient conditions for local exponential stability and global asymptotic stability (GAS) of the healthy state. Second, we provide sufficient conditions for existence, uniqueness, and  GAS of the endemic state. Building off of these results,} we provide a necessary, and sufficient, condition for the healthy state to be the unique equilibrium of our model. 
\seb{Third, we show that the endemic equilibrium of the SIWS model is worse than that of the networked 
SIS model without any infrastructure network, in the sense that at least one subpopulation has strictly larger infection proportion at the endemic equilibrium in the former model than that in the latter.} 
Fourth, we study an observability problem, and, assuming that the measurements of the sickness-levels of the human contact network are available, 
provide a necessary and sufficient condition for estimation
of 
the pathogen levels in the infrastructure network. 
\sebcancel{Furthermore, we provide another sufficient, but not necessary, condition for estimation of pathogen levels in the infrastructure network. 
By leveraging the 
sufficient condition
we finally provide insights in to how  the measurement matrix could be designed so that 
the system is locally weakly observable}.

\end{abstract}


\section{Introduction}
The spread of diseases \sebcancel{has} been a prominent feature of human civilization. The devastation that epidemics can 
\sebcancel{bring} worldwide, both from loss of life, and, less importantly, from hindrance to economic activity, has been brought into stark relief by the ongoing Covid-19 crisis. Consequently, understanding the causes of spread of diseases, and, as a result, possibly mitigating (or eradicating) the spread have been questions of longstanding interest for the scientific community. The earliest work in this area can be traced back to \cite{bernoulli1760essai}. In recent times, modeling and analysis  of spreading processes has attracted the attention of researchers across a wide spectrum ranging from mathematical epidemiology  \cite{bernoulli1760essai,hethcote2000mathematics} and physics \cite{van2009virus} to  the social sciences \cite{easley2010networks}.

Various models have been proposed in the literature for studying spreading processes, and, in particular, epidemics, viz. susceptible-exposed-infected-recovered (SEIR), susceptible-infected-susceptible (SIS), susceptible-infected-recovered (SIR), susceptible-infected (SI), and susceptible-infected-recovered-infected (SIRI). An overview of these models is provided in \cite{mei2017dynamics}, and \cite{pagliara18}, respectively. The present paper relies on the SIS model, which was first introduced in \cite{kermack1932contributions}. In an SIS model, an agent (resp. node), which can be interpreted as either an individual or, equivalently, a community, is either in the infected state or in the susceptible state. Assuming there is a non-trivial disease-spread in a population, an agent that is in the susceptible state, as a consequence of interactions with its neighbors, and depending on its infection rate, transitions to the infected state; an agent that is in the infected state recovers from the infections based on its healing rate. 

SIS networked models have been studied extensively in the literature, both for continuous-time and discrete-time settings; see, for instance, \cite{yorke2,van2009virus,khanafer2016stability,liu2019analysis} and \cite{peng2010epidemic,ahn2013global,pare2018analysis}, respectively. However, \cite{yorke2,van2009virus,khanafer2016stability,ahn2013global,pare2018analysis} only account for time-invariant interconnections among the agents. Overcoming this drawback,  SIS networked models with time-varying topologies have been developed and analysed  in \cite{prakash2010virus,rami2013stability,ogura2016stability}.  
Nonetheless, the bulk of the existing literature on SIS models factors in \emph{only} person-to-person interaction. However, diseases can spread also through other medium, such as water \cite{Vermeulen2015} (or infected surfaces, e.g., in hospitals \cite{weinstein2004contamination}, public transit vehicles \cite{hertzberg2018behaviors}, etc).  Water-borne 
pathogens could spread through infrastructure networks,  water distribution systems (e.g., rivers, groundwater, and reservoirs) \cite{Kough2015}. 
Moreover, while water quality issues are very prevalent in developing countries with less advanced plumbing and sewage infrastructure, such issues occasionally affect more prosperous countries as well. Notably, Sweden has had a number of water contamination incidents which have affected thousands of residents. For example, in \"{O}stersund in Northern Sweden,  approximately $27,000$ people ($\sim$45\% of the population) became ill and had a water-boil order for over two months as the result of Cryptosporidium contamination of the drinking water \cite{widerstrom2014large}. Hence, there is a need for SIS networked models that \emph{also} account for the spread of diseases through water distribution networks.

Based on the aforementioned motivation,  the so-called  Susceptible-Infected-Water-Recovered (SIWR) model had been proposed in \cite{tien2010multiple, tien2011herald,robertson2013heterogeneity} by adding a water compartment to the classical SIR model. More recently, a variant of the SIS model called the  Susceptible-Infected-Water-Susceptible (SIWS) model 
has been recently developed in \cite{liu2019networked}, and a multi-virus single resource SIWS model in \cite{axel2020TAC}. The paper \cite{liu2019networked} provides sufficient conditions for GAS of the healthy state (see \cite[Theorem~1]{liu2019networked}), but it does not provide any theoretical guarantees regarding endemic behavior. More recently,  sufficient conditions for GAS of the healthy state, and also for the existence, uniqueness, and GAS of the endemic state have been provided in \cite{axel2020TAC}; see \cite[Theorem~2]{axel2020TAC}, and \cite[Theorem~3]{axel2020TAC}, respectively. However, both \cite{liu2019networked} and \cite{axel2020TAC} consider only the presence of a \emph{single} resource. Notice that if there are \emph{multiple} water resources being accessed by the population, then the spread of virus could be due to not only node-to-node interaction and node-to-resource interactions, but also due to \emph{resource-to-resource} interaction. The present paper aims to develop such a model (called the layered networked SIWS model), and provide an in-depth analysis of its various equilibria viz.  existence, uniqueness, and stability. \seb{Based on the aforementioned analysis, we would also focus on understanding the effect on the endemic level of the population nodes in the presence of shared resource(s) as compared to the absence of the same.}

While the discussion insofar has been centered around modeling and analysis, another pressing challenge that health administration officials face is to 
estimate the contamination levels in the infrastructure network. In particular, for large-scale infrastructure networks (as is the case with modern societies), it is not economically viable to install sensors 
everywhere.
However, by employing system-theoretic notions such as observability\footnote{A system has the property of observability, if, given a series of output measurements, the initial state of the system can be uniquely determined.}, one could address the aforementioned challenge by 
deploying
as few sensors as possible. One of the earliest works in this direction is \cite{alaeddini2016optimal}, where the problem of \emph{which} subset of nodes in a network should be measured so as to improve observability of a SIS network is addressed; the condition therein involves checking the determinant of the inverse of the observability Grammian. Inspired by the work in \cite{alaeddini2016optimal},  we aim to address the following question: under what conditions can we 
estimate the contamination levels in the infrastructure network by only measuring the infection levels of individuals in the population? Furthermore, given knowledge of such conditions, can we glean any insights into how the measurement matrix might be designed so that the contamination levels in the infrastructure network can be recovered purely by  measuring the infection levels of individuals in the population?
The key theoretical tool that we would be using to answer these questions is the notion of local weak observability of non-linear systems\footnote{We say that  two initial states
are \emph{indistinguishable} if the corresponding outputs 
are equal for all time instants. A system is locally weakly observable if one can instantaneously distinguish each initial state from its neighbors~\cite{hermann1977nonlinear}. 
}. \seb{}

{\em Paper Contributions:}
 \sebcancel{For the layered networked SIWS model that accounts for the presence of multiple resources, our main contributions} are as follows:
\begin{enumerate}[label=(\roman*)]
    \item We 
    identify     conditions such that regardless of whether or not an agent (resp. infrastructure resource) is 
    infected
    or healthy, the model converges to the healthy state, i.e., conditions for global asymptotic stability (GAS) of the healthy state; see Theorem~\ref{thm:GAS}.
    \item We provide  conditions that guarantee the existence, uniqueness, and GAS of the endemic equilibrium; see Theorem~\ref{thm:equi}.
    \item \seb{We show that the endemic equilibrium in the population nodes for the layered networked SIWS model is greater than or equal to the endemic equilibrium of the population nodes in the networked SIS model, with at least one of the population nodes in the former having a strictly greater endemic level than  in the latter; see Proposition~\ref{prop:endemic_largerwithres}.}
    \item  Assuming all nodes in the human contact network are initially healthy, we provide a necessary and sufficient condition 
    for
    local weak observability
    of the layered networked SIWS model; see Theorem~\ref{thm:obs_x0=0}. 
\end{enumerate}

\sebcancel{Additionally, we also have the following auxiliary contributions:   a necessary, and sufficient, condition for the healthy state to be the unique equilibrium of 
    the
    model; see Corollary~\ref{cor:unique:healthystate}.   A sufficient (but not necessary) condition for local weak observability of the layered networked SIWS model, and, based off of this sufficient condition,
    we present a design of the observability matrix that results in the layered networked SIWS model being locally weakly observable; see Proposition~\ref{prop:sensor:placement} and Corollary~\ref{cor:sensor:placement:algo}, respectively.}\\
\indent  A preliminary version of this paper appeared in \cite{cdc_water}. \sebcancel{The present paper involves a more comprehensive treatment by providing theoretical guarantees for the endemic behavior, 
novel sufficient conditions for local weak observability, complete proofs of all assertions, and, finally, an in-depth set of simulations.} 

{\em Paper Organization:}
The paper unfolds as follows. We conclude the present section by collecting all the notation 
used in the 
rest of the paper.
The layered networked SIWS model is developed in Section~\ref{sect:model}, where, we subsequently, also state the problems of interest. The analysis of the various equilibria of the model, namely stability of the healthy state and existence, uniqueness, and stability of the endemic state, is given in Section~\ref{sect:analysis}. 
The observability problem
is studied in Section~\ref{sec:obsv}. Simulations illustrating our theoretical findings are provided in Section~\ref{sec:simulations}.  Finally, some concluding remarks, together with some research directions of possible interest to the wider community, are provided in Section~\ref{sec:conclusion}. 

{\em Notation:}
For any positive integer $n$, we use $[n]$ to denote the set $\{1,2,\ldots,n\}$.
The $i$th entry of a vector $x$ will be denoted by $x_i$.
We use $\0$ and $\1$ to denote the vectors whose entries all equal $0$ and $1$, respectively,
and use $I$ to denote the identity matrix.
For any vector $x\in\R^n$, we use ${\diag}(x)$ to denote the $n\times n$ diagonal matrix whose $i$th diagonal entry equals $x_i$.
For any two sets $\scr{A}$ and $\scr{B}$,
we use $\scr{A}\setminus \scr{B}$ to denote the set of elements in $\scr{A}$ but not in $\scr{B}$.
For any two real vectors $a,b\in\R^n$, we write $a\geq b$ if
$a_{i}\geq b_{i}$ for all $i\in[n]$,
$a>b$ if $a\geq b$ and $a\neq b$, and $a \gg b$ if $a_{i}> b_{i}$ for all $i\in[n]$.
For a square matrix $M$, we use 
$\sigma(M)$ to denote the spectrum of $M$, 
use $\rho(M)$ to denote the spectral radius of $M$, 
and $s(M)$ to denote the largest real part among the eigenvalues of $M$, i.e., $s(M) = \max \left\{{\rm Re}(\lambda)\ : \ \lambda\in\sigma(M)\right\}$. Given a matrix $A$, $A \prec 0$ (resp. $ A\preccurlyeq 0 $) indicates that $A$ is negative definite (resp. negative semidefinite), whereas $A \succ 0$ (resp. $ A\succcurlyeq 0 $) indicates that $A$ is positive definite (resp. positive semidefinite).


\section{The Model} \label{sect:model}
\begin{figure}
    \centering
    \begin{overpic}[width=0.9\columnwidth]{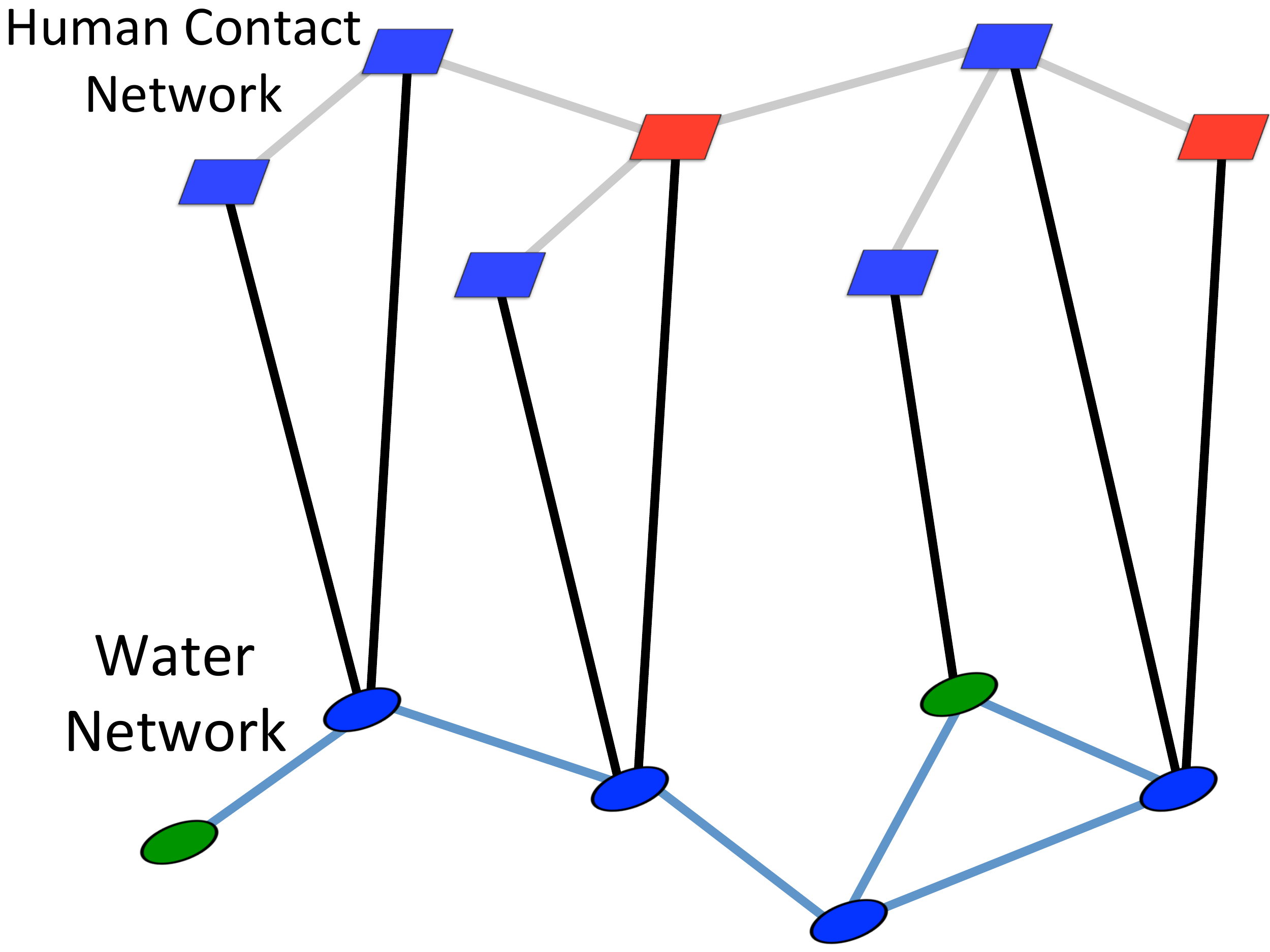}
    \put(-3.5,72){\colorbox{white}{\parbox{0.75\linewidth}
     \large 
    Human Contact
     }}
    \put(2.75,67){\colorbox{white}{\parbox{0.75\linewidth}
     \large 
    Network
     }}
     \put(-1.9,22){\colorbox{white}{\parbox{0.75\linewidth}
     \large 
    Infrastructure
     }}
     \put(3,17){\colorbox{white}{\parbox{0.75\linewidth}
     \large 
     Network
     }}
    \end{overpic}
    \caption{Multi-layered SIWS model: The disease (depicted by red) spreads between household nodes (squares) and the pathogen (green) spreads through 
    infrastructure  network nodes (circles). Blue indicates healthy. The model permits transmission from the 
    infrastructure network to the human contact network, vice versa, and not necessarily symmetrically.}
    \label{fig:siws}
\end{figure}
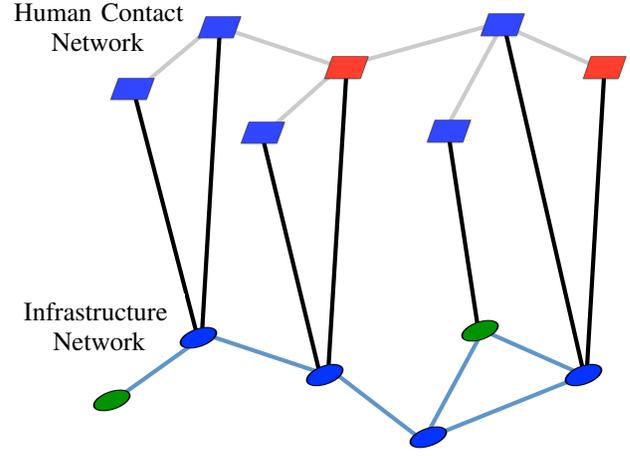

In this section, we develop a distributed continuous-time pathogen model. This model will be hereafter referred to as \emph{the layered networked SIWS model}; \sebcancel{see Figure~\ref{fig:siws}}.



\subsection{The layered networked SIWS model}
Consider a pathogen spreading over a two-layer network consisting of $n>1$ groups of individuals and $m>1$ infrastructure compartments. The individuals in a group could 
become
contaminated as a consequence of their interactions with other infected individuals and/or as a consequence of their interactions with infected infrastructure compartments.


We denote by $I_i(t)$ and $S_i(t)$ the number of infected and susceptible individuals, respectively, in group $i$ at time $t\ge 0$. We
denote by $N_i$ the total number of individuals in group $i$, and 
assume that $N_i$ does not change over time, i.e., $S_i(t)+I_i(t)=N_i$ for all $i\in [n]$ and $t\ge 0$, This assumption implies that the birth and death rates for each group are equal. Thus, it simplifies the model. 
The healing rate of each group~$i$ is denoted by $\gamma_i$,
the birth rate  by $\mu_i$, the death rate by $\bar \mu_i$ (which equals $\mu_i$), the person-to-person infection rates by $a_{ij}$ 
and the infrastructure-to-person infection rates by $a_{ij}^w$. \sebcancel{In the rest of this paper, we will assume that all of the aforementioned  parameters are nonnegative}.
We assume that the individuals are susceptible at birth regardless of whether (or not) their parents are infected.
The evolution of the numbers of infected and susceptible individuals in each group $i$ is, consistent with the ideas in  \cite{FallMMNP07,water}, as follows:

\begin{eqnarray}
\dot S_i(t) &=&  \mu_i N_i - \bar\mu_i S_i(t) + \gamma_iI_i(t) - \textstyle \sum_{j=1}^n a_{ij} \frac{S_i(t)}{N_i} I_j(t) \nonumber\\ 
 && - \textstyle \sum_{j=1}^m a_{ij}^w w_j(t)S_i(t) \nonumber
 \\
          &=&  (\mu_i + \gamma_i )I_i(t) - \textstyle \sum_{j=1}^n a_{ij} \frac{S_i(t)}{N_i} I_j(t) \nonumber\\ 
 && - \textstyle \sum_{j=1}^m a_{ij}^w w_j(t)S_i(t), \label{xxx1} 
 \\ \nonumber
 \dot I_i(t) &=& - \gamma_iI_i(t) - \bar\mu_i I_i(t) + \textstyle\sum_{j=1}^n a_{ij} \frac{S_i(t)}{N_i} I_j(t) \nonumber
  \end{eqnarray}
  \begin{eqnarray}
  && + \textstyle\sum_{j=1}^m \alpha_{ij}^w w_j(t)S_i(t) \nonumber \\
          &=& (-\gamma_i-\mu_i) I_i(t) + \textstyle\sum_{j=1}^n a_{ij} \frac{S_i(t)}{N_i} I_j(t) \nonumber \\
 && + \textstyle\sum_{j=1}^m a_{ij}^w w_j(t)S_i(t), \label{xxx2}
 \end{eqnarray}
 

\noindent
where $w_j(t)$ denotes the pathogen concentration in the $j$th infrastructure compartment and 
evolves as 
\begin{equation}
 \dot w_j =  - \delta_j^w w_j + \textstyle\sum_{k=1}^n \zeta_{jk}^w I_k +
\textstyle\sum_{k = 1 }^m
    \alpha_{kj} w_k - w_j \sum_{k = 1 }^m 
    \alpha_{jk},
\label{www}\end{equation}
 

\noindent
where $\delta^w_j$ denotes the decay rate of the pathogen, 
$\zeta_{jk}^w$ denotes the person-infrastructure contact rate of group $k$ to infrastructure node $j$, 
and 
$\alpha_{kj}$ represents the flow of the pathogen from node $k$ to node $j$ in the infrastructure network.
It is 
clear 
from \eqref{xxx1} and \eqref{xxx2}, that $\dot S_i(t) + \dot I_i(t) = 0$, which is 
consistent
with our assumption that $N_i$ is a constant. 

We simplify the model further by defining the fraction of infected individuals in each group $i$ 
as
$$x_i(t) = \frac{I_i(t)}{N_i}. $$
 

\noindent
By defining the following parameters
$$\delta_i=\gamma_i+\mu_i, \;\;\; \beta_{ij} = a_{ij} \frac{N_j}{N_i}, \;\;\; \beta_{ij}^w = 
N_i a_{ij}^w
, \;\;\; c_{jk}^w = \zeta_{jk}^w / N_k
$$
and 
from \rep{xxx1}, 
\rep{xxx2}, and \rep{www}, it follows that
 \vspace{-0.5ex}
\begin{align}
    \dot x_i &=-\delta_i x_i+(1-x_i)\left(\textstyle\sum_{j=1}^n \beta_{ij}x_j+ \textstyle\sum_{j=1}^m\beta^w_{ij} w_j\right), \label{xi} \\
    \dot w_j &= -\delta^w_{j} w_j  + \textstyle\sum_{k = 1 }^m
    \alpha_{kj} w_k - w_j \textstyle\sum_{k = 1 }^m 
    \alpha_{jk} + \textstyle\sum_{k=1}^n c^w_{jk} x_k. \label{wj}
\end{align}




\noindent
Note that, we also allow for the healing rate of an infrastructure compartment $j$, $\delta^w_j$, to be zero. 

The model from 
\eqref{xi}-\eqref{wj} in vector form becomes:
\begin{align}
    \dot x &=(B - XB - D)x + (I-X)B_w w \label{x} \\
    \dot w &= -D_w w  + A_w w + C_w x, \label{w}
\end{align}
 

\noindent
where $B = [\beta_{ij}]_{n\times n}$, $X={\diag}(x)$, $B_w = [\beta^w_{ij}]_{n\times m}$, $A_w$ has off-diagonal entries equal to $\alpha_{kj}$ and diagonal entries equal to $-\sum_{k}\alpha_{kj}$, and $C_w = [c^w_{jk}]_{m\times n}$. Therefore, the columns of $A_w$ sum to zero. 

\noindent System~\eqref{x}-\eqref{w} 
could be written more compactly
using
%
\begin{align}
\nonumber
 z(t) 
 &\coloneqq
 \begin{bmatrix}
 x(t)\\
 w(t)
 \end{bmatrix}
 ,
 \,\,
 X(z(t))  
 \coloneqq
 \begin{bmatrix}
 \diag(x(t)) & 0 \\
 0 & 0 
 \end{bmatrix},
 \\
 \normalsize B_f 
 &\coloneqq
 \begin{bmatrix}
 B & B_w \\
 C_w & A_w - \diag (A_w) 
 \end{bmatrix}
 ,
 \text{ and}\label{eq:z}
 \\
 D_f 
 &\coloneqq
 \begin{bmatrix}
 D & 0 \\
 0 & D_w  - \diag (A_w)
 \end{bmatrix}
\nonumber
 .
\end{align}
\noindent With the new notations in place, 
\eqref{x}-\eqref{w} can be rewritten as:
\begin{equation} \label{eq:full}
    \dot{z} = \big{(} - D_f + (I - X(z)) B_f \big{)} z.
\end{equation}

\begin{rem}\label{rem:axel:TAC}
We highlight how the model considered in the present paper is connected with similar models in the existing literature
\begin{enumerate}[label=\roman*)]
    \item If $m=1$, 
 \eqref{eq:full} coincides with the model in \cite{liu2019networked}, and 
 with the multi-virus model  in \cite{axel2020TAC}, when the latter is particularized for the single-virus case.
 \item If $w_j(t)=0$ for all $t$ and all $j\in [m]$, or equivalently, there is no coupled infrastructure network, 
\eqref{eq:full} reduces to the regular networked SIS model in \cite{survey}.
\end{enumerate}
\end{rem}

\subsection{Problem Statements}\label{sec:pb:stmnt}
In the sequel, for the model in \sebcancel{\eqref{eq:full},} 
we will be interested in addressing the following problems:
\begin{enumerate}[label=(\roman*)]
\item \label{q1} \sebcancel{Identify a condition such that $z(t)$ converges asymptotically to the healthy state, i.e.,  $z = \textbf{0}$. }
\item \label{q2}
\sebcancel{ Under what conditions does there exist an endemic equilibrium  $\hat{z} > \textbf{0}$, and under such conditions, does the system converge asymptotically to $\hat{z}$ from any non-zero initial condition?}

 \item \label{q3:bis}
 \sebcancel{Let 
$\hat z = \begin{bmatrix} \hat x & \hat w
 \end{bmatrix}^\top$,    
 where $\hat x$ (resp. $\hat w$) denotes the endemic equilibrium of the population nodes (resp. the shared resources).
 Let  $\tilde{x}$ denote the unique endemic equilibrium of the SIS model without a shared resource. What is the relation between $\hat x$ and $\tilde x$? }
\item \label{q5}
\sebcancel{Identify a necessary, and sufficient, condition such that, given $x(t)$,  $z(0)$ can be uniquely recovered.}

\end{enumerate}

\subsection{Positivity Assumptions}
We impose the following assumptions on the parameters.

\begin{assum} \label{assum:sys:params}
Suppose that $\delta_i>0$ for all $i\in[n]$, $\delta^w_{j}+ \sum_{k}\alpha_{kj}>0$ 
for all $j\in[m]$, $\beta_{ij}\ge 0$ for all $i,j\in[n]$, 
and
$\beta_{ij}>0$ whenever group $j$ is a neighbor of group $i$.
\label{para}
\end{assum}



 


Since each $x_i$ represents the fraction of infected individuals
in group $i$, it is immediate that the initial value of $x_i$ is in $[0,1]$,
because otherwise the value of $x_i$ will lack physical meaning for the epidemic model considered here.
Similarly, it is also natural to assume that the initial value of $w_j$ \sebcancel{(measured, for instance, in milligrams per litre)} is nonnegative. 
Hence, we can
restrict our analysis to the set:
\begin{equation}\label{eq:set:D}
\mathcal{D} \coloneqq \{y(t): x(t) \in [0,1]^n, w(t) \in [0, \infty)^m\}.
\end{equation}
The following lemma establishes that the set $\mathcal D$ is positively invariant.

\begin{lem} \label{lem:z_invariantset}
Suppose that Assumption \ref{para} holds. Suppose that $x_i(0)\in[0,1]$ for all $i\in[n]$ and $w_j(0)\ge 0$ for all $j\in[m]$. Then, $x_i(t)\in[0,1]$ for all $i\in[n]$ and $w_j(t)\ge 0$ for all $j\in[m]$, for all $t\ge 0$.
\label{box}
\end{lem}

{\em Proof:}
Suppose that at some time $\tau$, $x_i(\tau)\in[0,1]$ for all $i\in[n]$ and $w_j(\tau)\geq 0$ for all $j\in[m]$.
First consider any index $j\in[m]$. If $w_j(\tau)=0$, then from \rep{wj} and Assumption~\ref{para}, $\dot w_j(\tau)\ge 0$.
Therefore $w_j(t)\ge 0$ for all  $t\ge \tau$.

Now consider any index $i\in[n]$.
If $x_i(\tau)=0$, then from \rep{xi} and Assumption \ref{para}, $\dot x_i(\tau)\ge 0$.
If $x_i(\tau)=1$, then again from \rep{xi} and Assumption \ref{para}, $\dot x_i(\tau)< 0$.
Therefore, $x_i(t)$ will be in $[0,1]$ for all times $t\ge \tau$.

Since the above arguments hold for any $i\in[n]$ and any $j\in[m]$,  we have that $x_i(t)\in[0,1]$ for
all $i\in[n]$ and $w_j(t)\geq 0$ for
all $j\in[m]$, $t\ge \tau$.
Since it is assumed that $x_i(0)\in[0,1]$ for all $i\in[n]$ and $w_j(0)\ge 0$ for all $j\in[m]$,
the lemma follows by setting $\tau = 0$.~$\square$

\section{Stability analysis of the equilibria}
\label{sect:analysis}

In this section\sebcancel{,} we analyze the equilibria of the proposed model and their stability both locally and globally.

\subsection{Local Stability of the Healthy State}

Consider $(\tilde x,\tilde w)$, an equilibrium of \eqref{x}-\eqref{w}.
The Jacobian matrix of the equilibrium, denoted by $J(\tilde x,\tilde w)$, is
\begin{equation}\label{jacob}
J(\tilde x,\tilde w) = 
\begin{bmatrix}
B-\tilde X B - D - F_1 - F_2 & (I-\tilde X)B_w\\
C_w & -D_w + A_w 
\end{bmatrix},
\end{equation}
where 
$\tilde X, F_1, F_2$ are diagonal matrices given by 
\begin{align}
\tilde X &= \diag\big(\tilde x_1, \tilde x_2, \cdots, \tilde x_n\big), \label{xtilde}\\
F_1 &= \diag\big(\textstyle\sum_{j=1}^n \beta_{1j}\tilde x_j, \textstyle\sum_{j=1}^n \beta_{2j}\tilde x_j, \cdots, \textstyle\sum_{j=1}^n \beta_{nj}\tilde x_j\big), \label{F1} \\
F_2 &= \diag\big(\textstyle\sum_{j=1}^n \beta_{1j}^w\tilde w_j, \textstyle\sum_{j=1}^n \beta_{2j}^w\tilde w_j, \cdots, \textstyle\sum_{j=1}^n \beta_{nj}^w\tilde w_j \big) \label{F2}.
\end{align}
In the case when $\tilde x = \0$ and $\tilde w = \0$,
i.e., at the healthy state \sebcancel{(also referred to as the disease-free equilibrium)},
\begin{equation*}\label{zerojacob}
J(\0,\0) = 
\begin{bmatrix}
B-D &  B_w \cr
C_w &  A_w - D_w
\end{bmatrix} =B_f-D_f.
\end{equation*}
If either $B_w = 0$ or $C_w = 0$, 
i.e., the pathogen does not affect the population or humans can not contaminate the infrastructure network by using it, 
we have the following result. 

\begin{prop} \label{prop:healthystate:locallyexpo:stable}
If $s(B-D) <0$, $s(A_w-D_w ) <0$, and $B_w = 0$ or $C_w = 0$, then the healthy state $(\0,\0)$ of \eqref{x}-\eqref{w} is locally exponentially stable.
\end{prop}

{\em Proof:}
If $B_w = 0$ or $C_w = 0$ then $J(\0,\0)$ is a triangular matrix (lower or upper, respectively), and therefore the spectrum of the matrix is equal to the union of the spectrum of the two block matrices on the diagonal. Consequently, if $s_1(B-D) <0$ and $s_1(A_w-D_w) <0$ then $J(\0,\0)$ is Hurwitz and by Lyapunov's indirect method \cite{khalil} the healthy state $(\0,\0)$ of \eqref{x}-\eqref{w} is locally exponentially stable.~$\square$

For nonzero $B_w$ and $C_w$, 
we have the following result. 

\begin{prop}\label{jjj}
Let Assumption \ref{para} hold. If $\rho(D_f^{-1}B_f)<1$ and $B_f$ is irreducible, then the healthy state $(\0,\0)$ of \eqref{x}-\eqref{w} is locally exponentially stable.
\end{prop}
\seb{\textit{Proof}: See Appendix.~$\square$}




\subsection{Global Stability of the Healthy State}

To state our first main result, we need the following concept. 
Consider an autonomous system
$\dot x(t)  = f(x(t))$, 
where $f: \scr{D}\rightarrow\R^n$ is a locally Lipschitz map from a domain
$\scr{D}\subset\R^n$ into $\R^n$.
Let $\tilde x$ be an equilibrium of the system and $\scr{E}\subset\scr{D}$ be a domain containing $\tilde x$.
The equilibrium $\tilde x$ is called asymptotically stable with the domain of attraction $\scr{E}$ if
for any $x(0)\in\scr{E}$, there holds
$\lim_{t\rightarrow\infty}x(t) = \tilde x$.

The global stability of the healthy state is characterized by the following theorem.

\begin{thm}
Let Assumption \ref{para} hold.
If $\rho(D_f^{-1}B_f)\leq 1$  and $B_f$ is irreducible, 
then the healthy state 
of \eqref{x}-\eqref{w} is asymptotically stable with the domain of attraction $\mathcal D$, with $\mathcal D$ 
given in~\eqref{eq:set:D}. 
\label{thm:GAS}\end{thm}
\textit{Proof:} See Appendix.~$\square$\\
Theorem~\ref{thm:GAS} addresses     Question~\ref{q1} in Section~\ref{sec:pb:stmnt}.

\subsection{Reproduction Number}

In epidemiology the reproduction number, $R_0$, is the average number of people that become  infected from one infected individual. If $R_0>1$ the disease will lead to an outbreak; if $R_0\leq 1$ the disease will die out. 
For the networked SIS model with no water  compartments, it has been shown that $\rho(D^{-1}B)$ is the reproduction number, and that if $\rho(D^{-1}B)\le 1$, the model will asymptotically converge to the healthy state for all initial conditions, and if $\rho(D^{-1}B)>1$, the model will asymptotically converge to a unique epidemic state for all initial conditions except for the healthy state \cite{FallMMNP07}. 

For the layered networked SIWS model \rep{x}-\rep{w}, Theorem~\ref{thm:GAS} implies that when $\rho(D_f^{-1}B_f)\le 1$, the model will asymptotically converge to the healthy state for all initial conditions, which implies that the healthy state is the unique equilibrium. 
We call $\rho(D_f^{-1}B_f)$ the basic reproduction number of the layered networked SIWS model \rep{x}-\rep{w}, and compare its value with that of the networked SIS model, $\rho(D^{-1}B)$, to illustrate the effect of the water distribution network. 
Note that 

\vspace{-2ex}\scriptsize
\begin{align*}
D_f^{-1}B_f &= 
\begin{bmatrix}
D^{-1}  &  0 \cr
0  &  (D_w - \diag(A_w))^{-1}
\end{bmatrix}\begin{bmatrix}
B  &  B_w \cr
C_w  &  A_w - \diag(A_w)
\end{bmatrix} \\
&= \begin{bmatrix}
D^{-1}B  &  D^{-1}B_w \cr
(D_w - \diag(A_w))^{-1} C_w  &  (D_w - \diag(A_w))^{-1} A_w - \diag(A_w)
\end{bmatrix}.
\end{align*}\normalsize
We need the following lemma.

\begin{lem}
{\cite[Lemma 2.6 ]{varga}}
Suppose that $N$ is an irreducible nonnegative matrix. If $M$ is a principal square submatrix of $N$, then $\rho(M)<\rho(N)$.
\label{prin}\end{lem}

Since $D_f^{-1}B_f$ is an irreducible nonnegative matrix by Assumption~\ref{para}, and since $D^{-1}B$ is a principal square submatrix of $D_f^{-1}B_f)$, from Lemma~\ref{prin} it follows that $\rho(D_f^{-1}B_f)>\rho(D^{-1}B)$. Therefore we have the following result. 

\begin{prop}\label{prop:reprod:number}
Suppose that Assumption~\ref{para} holds.
Then, the basic reproduction number of the layered networked SIWS model \rep{x}-\rep{w} is greater than that of the networked SIS model.
\end{prop}
\noindent 
Proposition~\ref{prop:reprod:number} implies that eradication of the disease in the population in itself \sebcancel{does not guarantee}
that the system is disease-free. That is, the presence of infrastructure network makes the system more vulnerable to SIS-type diseases than otherwise.

\subsection{Analysis of the endemic behavior} \label{sec:endemic:behavior}
It turns out that the condition in Proposition~\ref{prop:healthystate:locallyexpo:stable} being violated results in the instability of the healthy state $(\0,\0)$ of \eqref{x}-\eqref{w}, as we show in the following proposition.
\begin{prop}\label{prop:healthystate:unstable}
Suppose that $B_w = 0$ or $C_w = 0$. If $s(B-D) >0$ or $s(A_w-D_w ) >0$, then the healthy state $(\0,\0)$ of \eqref{x}-\eqref{w} is unstable.
\end{prop}

\textit{Proof:} Since by assumption, $B_w = 0$ (resp. $C_w = 0$), it follows that the Jacobian matrix of the equilibrium evaluated at the healthy state, i.e., $J(\0,\0)$, is a block lower triangular (resp. upper triangular)
matrix. Hence, the eigenvalues of 
$J(\0,\0)$ are same as those of matrices $B-D$ and $A_w-D_w$. Consequently, if $s(B-D) >0$ and/or $s(A_w-D_w ) >0$, then $s(J(\0,\0)) >0$. Hence, the healthy state $(\0,\0)$ of \eqref{x}-\eqref{w} is unstable.~$\square$\\

\seb{Simulations indicate the existence of an endemic \sebcancel{state (also referred to as the endemic equilibrium)} when the eigenvalue condition in Theorem~\ref{thm:GAS} is violated (see Figure~\ref{fig:assum1_endem} in Section~\ref{sec:simulations}), a rigorous result, however, remains missing. Therefore,}
we consider the following variant of Assumption~\ref{assum:sys:params}.



\begin{assum} \label{assum:base}
    Assume that $\delta_i > 0$, $\delta_j^w > 0$, $\beta_{ij} \geq 0$, $\beta_{ij}^w \geq 0$, and that, \sebcancel{for $i \neq j$}, $\alpha_{ij} \geq 0$, with $\alpha_{ii} = - \textstyle \sum_{j \neq i}^m \alpha_{ji}$.~$\blacksquare$
\end{assum}
\sebcancel{Assumption~\ref{assum:base} states that the system parameters, with the exception of the rate of flow of pathogen within a resource node, are nonnegative}.
\seb{It is easy to show that Assumption~\ref{assum:base} implies Assumption~\ref{assum:sys:params}, and, is, thus, more restrictive. Hence,  we only need  Assumption~\ref{assum:base} in the sequel. }

\begin{thm} \label{thm:equi}
Consider~\eqref{eq:full} under Assumption~\ref{assum:base}. Suppose that $B_f$ is irreducible and $\rho (D_f^{-1} B_f) > 1$. Then there exists a unique endemic equilibrium $\Tilde{z} \gg \textbf{0}$. Furthermore, $\Tilde{z}$ is asymptotically stable with the domain of attraction $\mathcal D \setminus \{\textbf{0}\}$, with $\mathcal D$
given in~\eqref{eq:set:D}.~$\blacksquare$
\end{thm}
\textit{Proof:} See Appendix.~$\square$

Theorem~\ref{thm:equi}  says that as long as the reproduction number of the layered SIWS network is greater than one, then, assuming that there is at least one node (population or infrastructure) that is infected initially, the spreading process converges to a unique proportion in each population node, and a unique 
infection level in each infrastructure node. Thus, Theorem~\ref{thm:equi} addresses Question~\ref{q2} in Section~\ref{sec:pb:stmnt}. Note that Theorem~\ref{thm:equi} improves upon \cite[Theorem~3]{axel2020TAC} since it also accounts for \emph{multiple} shared resources.

Combining Theorems~\ref{thm:GAS} and~\ref{thm:equi} yields a necessary, and sufficient, condition for the healthy state to be the unique equilibrium of 
\eqref{x}-\eqref{w}. Hence, we have the following result:
\begin{cor}\label{cor:unique:healthystate}
Consider the layered networked SIWS model in~\eqref{x}-\eqref{w} under Assumption~\ref{assum:base}. 
Suppose that $B_f$ is irreducible. Then the healthy state is the unique equilibrium in the domain $\mathcal D$ if, and only if, $\rho(D_f^{-1}B_f) \leq 1$.~$\blacksquare$
\end{cor}

Rewriting the condition in Corollary~\ref{cor:unique:healthystate} in view of \cite[Proposition~1]{liu2019analysis} tells us that insofar the linearized state matrix of system~\eqref{eq:full} (linearized around the healthy state) is Hurwitz, the healthy state is the \emph{only} equilibrium of system~\eqref{eq:full}.

A very pertinent question that could arise at this point is as follows: focusing solely on the population, is there a relation between the endemic equilibrium of the layered networked SIWS model, and that of the networked SIS model. In order to answer this, we recall the latter:
\begin{equation}
    \dot x =(B - XB - D)x.  \label{x_SIS} 
\end{equation}
In order to ensure that the model in~\eqref{x_SIS} is well-defined, we need to particularize Assumption~\ref{assum:base}, 
for the setting without  shared resource(s). This is given as follows:
\begin{assum} \label{assum:noshared}
Suppose that $\delta_i>0$ and $\beta_{ij} \geq 0$ for all $i, j \in [n]$. 
\end{assum}
Let  $\tilde{x}$ denote the unique endemic equilibrium of~~\eqref{x_SIS} and 
\begin{equation*} \scriptsize
\hat z = \begin{bmatrix} \hat x \\ \hat w
 \end{bmatrix}    
\end{equation*} 
denote the unique endemic equilibrium of~\eqref{eq:full}. With this notation
and Assumption~\ref{assum:noshared} in place, we present the following result.

\begin{prop}\label{prop:endemic_largerwithres}
 Consider~\eqref{eq:full} under Assumption~\ref{assum:base}, and~\eqref{x_SIS} under Assumption~\ref{assum:noshared}. Suppose that $B_f$ and $B$ are irreducible, and that $\rho(D_f^{-1}B_f)>1$, and $\rho(D^{-1}B)>1$. 
 Then $\hat{x} > \tilde{x}$. 
\end{prop}
\textit{Proof}: See Appendix.~$\square$\\

Proposition~\ref{prop:endemic_largerwithres} says that the endemic level in each of the population nodes for the layered networked SIWS model is greater than or equal to the endemic level of the population nodes in the absence of shared resource(s). As such, it addresses Question~\ref{q3:bis} in Section~\ref{sec:pb:stmnt}.



\section{Observability 
Problem
}\label{sec:obsv}

In this section, we aim to address the following question: 
Assuming enough sensors to detect waterborne pathogens are not available,
can 
measurements of the sickness level of people, or households, who drink the water be used to estimate the contamination level of the water, or the source of the contamination (the initial condition of the water pathogen levels). 
We introduce the following notation:
\begin{equation}\label{y}
    y = Cx,
\end{equation}
where $C \in \mathbb R^{q \times n}$ is a measurement matrix, with $q \in \mathbb Z_+$ denoting the number of measurements needed. The problem posed in Question~\ref{q5} 
could be re-written as follows: 
Given $B$, $D$, $A_w$, $B_w$, $C_w$, $D_w$, $C$, and measurements $y$, find conditions for when $w(0)$ can be recovered.

We derive conditions \sebcancel{such that, given measurements of sickness levels of households, it is possible to uniquely recover the initial state of the water pathogen levels. Towards this end, we}
appeal to the rank of the Jacobian of the Lie derivatives, and apply the results from  \cite{hermann1977nonlinear}.
Consequently, the Lie derivative calculations are as follows:
\begin{align*}
    y &= C x 
    \\
    \dot y &= C \dot x = C\big(\underbrace{(B-XB-D)}_{F_x}x 
              + \underbrace{(I - X)B_w}_{F_w}w\big) \\
    \ddot{y} &= C \ddot{x} = C \big(F_x \dot x 
            + F_w \dot w - \dot X (Bx + B_w w)\big) \\
    y^{(3)} &= C x^{(3)} = C \big(F_x \ddot x 
            + F_w \ddot w - \ddot X (Bx + B_w w)
    \\
              & \; \; \; \; \; \; \; \;\; \; \; \;\; \; \; \; \; \; \; \; \; \; \; \; \; \; \; \; \; \; \; \; \; \; \; \; \; \; \; \; \; \; \; \; \; \; \; \;   
              - 2\dot X (B\dot x + B_w \dot w) \big) \\
    & \hspace{1.25ex}  \vdots \\
    y^{(m+n)} &= C x^{(m+n)} = C \big(F_x x^{(m+n-1)} 
            + F_w w^{(m+n-1)} 
    \\
              & \;  \; \; \; \; \; \; \; \;\; \; \; \;\; \; \; \; \;
              - X^{(m+n-1)} (Bx + B_w w) - \cdots 
              \big) ,
\end{align*}
where $\dot x$ and $\dot w$ are defined in \eqref{x} and \eqref{w}, 
\begin{align*}
    \ddot w &= (A_w-D_w)\dot w + C_w \dot x, \\
    \dot X &= {\rm diag}(\dot x), \\
    \ddot X &= (\tilde B_{x}+\tilde B_w )(F_x - \tilde B_{xw}),
\end{align*}
with $\tilde B_{x} = {\rm diag}(Bx)$, and $\tilde B_{w} = {\rm diag}(B_w w)$.

We explore the case when we assume that all nodes in the human contact network are initially healthy, that is, $x(0) = \0$.
Therefore, we explore the Jacobian of the above Lie derivatives evaluated at $x(0) = \0$, called $\mathcal{O}$, \sebcancel{where $\mathcal{O}=$}

\vspace{-2ex}

\small
\begin{equation}\label{O}
    \begin{bmatrix}
      C & 0 \\
      C(\underbrace{F_{x_0} - \tilde B_w}_{X_x}) & C B_w \\
      C(\underbrace{ X_x^2  - \tilde B_w B -\tilde B}_{X_{xx}}) & 
      C(\underbrace{B_w F_{w_0} + X_x B_w - \tilde B_w B_w}_{W_{w}})
    \\
      C(F_{x_0}X_{xx} - \tilde D B - \tilde D_w
      &
      C(B_w F_{w_0}^2 +X_x W_w - 2 \tilde B_w^2 B_w  \\
      -2\tilde B_w^2 B   - 2\tilde B_w B X_x &  - \tilde B_w (B_w F_{w_0} + B B_w)
      \\
      -2\tilde B X_x - {\rm diag}(B B_w w) &  - 2 \tilde B W_w - \tilde D B_w + B_w C_w B_w)\\
      +B_w(F_{w_0}C_w + C_w X_x)
      )\\
      \vdots & \vdots\\
    \end{bmatrix},
\end{equation}
\normalsize
with $F_{x_0} = (B-D)$, $F_{w_0} = (A_w-D_w)$, 

\vspace{-2ex}

\small
\begin{align*}
    \tilde B &= {\rm diag}(BB_w w + B_w F_{w_0}w), \\
    \tilde D &= {\rm diag}(F_{x_0}B_w w), \\ 
    \tilde D_w &= {\rm diag}(B X_x B_w w - B^2 F_{w_0}w + B_w F_{w_0}^2 w + B_w C_w B_w w).
\end{align*}
\normalsize
\sebcancel{Furthermore, $\mathcal{O}$ has $q(n+m)$ rows, and $n+m$ columns.}

Therefore, from   \cite[Theorems 3.1 and 3.12]{hermann1977nonlinear}, and since the system is analytic,
we have the following theorem. 
\begin{thm}\label{thm:obs_x0=0}
The layered networked SIWS model in \eqref{x}-\eqref{w} with measurements in \eqref{y} is 
locally weakly observable at $x(0) = \0$ if and only if  
$\mathcal{O}$,
as defined in \eqref{O}, has full rank.
\end{thm}



Observe that Theorem~\ref{thm:obs_x0=0} provides a necessary and sufficient condition for checking whether (or not) the layered networked SIWS model is locally weakly observable at $x(0) =\mathbf{0}$, \seb{and thus answers Question~\ref{q5} in Section~\ref{sec:pb:stmnt}}. However, the condition therein involves checking the rank of 
the $\mathcal{O}$ matrix, which in turn, involves 
 too many computations, \sebcancel{since $\mathcal{O}$ has $q(n+m)$ rows.}.
This drawback motivates us to seek a simpler, 
easier to check, sufficient condition for the layered networked SIWS model to be locally weakly observable at $x(0) =\mathbf{0}$, and is presented next.
\begin{prop}\label{prop:sensor:placement}
Suppose that the matrices $C$ and $CB_w$ have full column rank. Then  the layered networked SIWS model in \eqref{x}-\eqref{w} with measurements in \eqref{y} is locally weakly
observable at $x(0) = \0$.
\end{prop}
\textit{Proof:} Define the matrix 
\begin{equation} \label{matrix F}
\mathcal F: =
\begin{bmatrix}
C &0\\
CX_x & CB_w
\end{bmatrix}
\end{equation}
\sebcancel{Observe that $\mathcal F$ is a block lower triangular matrix. Hence, $\rank(\mathcal F) \geq \rank(C) + \rank(CB_w)$. By assumption, matrices $C$ and $CB_w$ have full column rank, which implies that $\rank(\mathcal F) \geq n + m$. Observe also that the total number of columns in $\mathcal F$ equals $n + m$. Therefore, it follows that  $\rank(\mathcal F) \leq n+m$. Hence, $\rank (\mathcal F) = n+m$, i.e., $\mathcal F$ has full column rank.}
Now note that $\mathcal F$ is a submatrix of $\mathcal O$, that has the \emph{same} number of columns as $\mathcal O$. 
Also observe that adding more rows to $\mathcal F$ does not lead to matrix $\mathcal F$ becoming rank deficient. This implies that matrix $\mathcal O$ has full column rank, and therefore, from Theorem~\ref{thm:obs_x0=0}, we conclude that the layered networked SIWS model in \eqref{x}-\eqref{w} with measurements in \eqref{y} is locally weakly
observable at $x(0) = \0$.~$\square$\\
Given that both Proposition~\ref{prop:sensor:placement}  and Theorem~\ref{thm:obs_x0=0} provide sufficient conditions for local weak observability, it is natural to ask how the two conditions are related. 
The following remark addresses this question.
\begin{rem}\label{rem:number:of:measurements} 
 Proposition~\ref{prop:sensor:placement} implies Theorem~\ref{thm:obs_x0=0}. The converse, however, is not true.
To see this, consider the following example:
Let $n=2, m=2$. With $C = I$, $D = I$, $D_w = I$,
\begin{align*}
    &\,\,\,
    B=
    \begin{bmatrix}
    1 & 1 \\
    1 & 2
    \end{bmatrix}
    \,
    , \, \,
    B_w=
    \begin{bmatrix}
    1 & 0 \\
    1 & 0
    \end{bmatrix}, \\
    &C_w=
    \begin{bmatrix}
    1 & 1 \\
    1 & 1
    \end{bmatrix}
    \,
    , \, \,
    A_w=
    \begin{bmatrix}
    1 & 1 \\
    1 & 1
    \end{bmatrix}
    ,
\end{align*}
it is clear that $C B_w$ does not have full column rank, so the conditions for Proposition~\ref{prop:sensor:placement} are not met. 
However, allowing $w = (w_1, w_2)$ to be free, we obtain 
\begin{align*}
    &C W_w=
    \begin{bmatrix}
    1-2 w_1 & 1 \\
    2-2 w_1 & 1
    \end{bmatrix}
    .
\end{align*}
Therefore, independent of the value of $w$, the rightmost column of $\mathcal{O}$ is linearly independent of the other three columns of $\mathcal{O}$, that is,
$\mathcal{O}$ has full column rank. Thus, the condition in Theorem~\ref{thm:obs_x0=0} is met.~$\blacksquare$
\end{rem}
We now highlight an interesting consequence of Proposition~\ref{prop:sensor:placement}. 
\begin{cor}\label{cor:sensor:placement:algo}
Let $n \geq m$. If $C = I_{n \times n}$ and $B_w$ has full column rank, then the layered networked SIWS model in \eqref{x}-\eqref{w} with measurements in \eqref{y} is locally weakly
observable at $x(0) = \0$.
\end{cor}
\textit{Proof:} Suppose that, by assumption, $C=I_{n  \times n}$. Consequently,  $\rank(C) =n$. Moreover, $CB_w =B_w$, and hence $\rank(CB_w) = \rank(B_w)$. Since, by assumption, $\rank(B_w) =m$, it follows that the conditions in Proposition~\ref{prop:sensor:placement} are satisfied, and hence the result follows.~$\square$\\
Observe that the result in Corollary~\ref{cor:sensor:placement:algo} could potentially inform sensor placement (in the population) strategies for detecting contamination levels of water resources in the layered networked SIWS model. 

The condition in Proposition~\ref{prop:sensor:placement} requires the observation matrix, $C$, to have full column rank. This implies that $q \geq n$. That is, the number of observations should 
at least 
be equivalent to
the size of the population. Clearly, for a large population,
this condition is quite restrictive. Hence, it is perhaps more appealing to establish conditions for local weak observability under partial measurements, which is beyond the scope of the present paper.

\section{Simulations}\label{sec:simulations}

\begin{figure}[t!]
	\centering
    \begin{minipage}[c]{0.55 \columnwidth}
        \includegraphics[width=1.2\columnwidth]{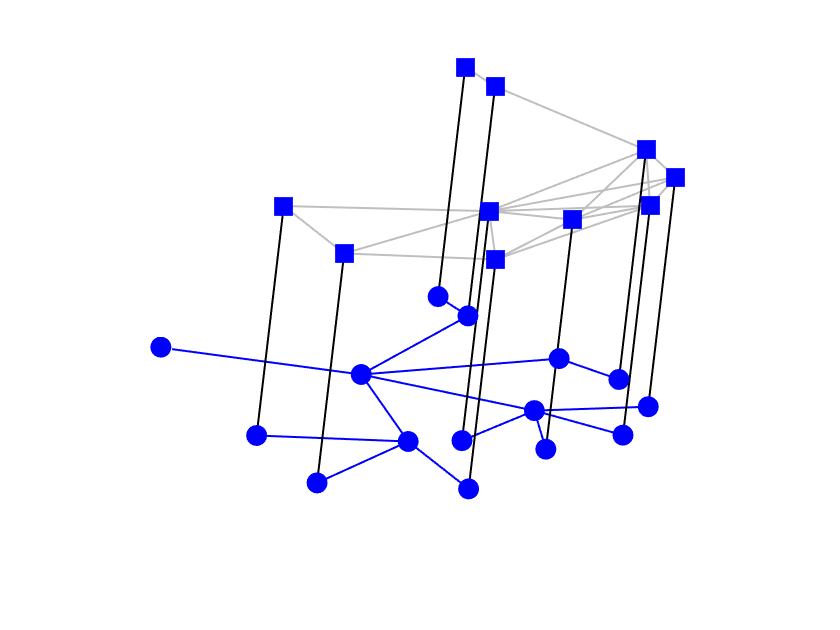}
    \end{minipage}\hfill
    \begin{minipage}[c]{0.4\columnwidth}
        \caption{The contact network of population and resource nodes used for simulations, represented by squares and circles, respectively.} 
        \label{fig:network}
    \end{minipage}
\end{figure}



\begin{figure}[t!]
	\centering
    \begin{subfigure}[b]{.49\columnwidth}
   \includegraphics[width=\columnwidth]{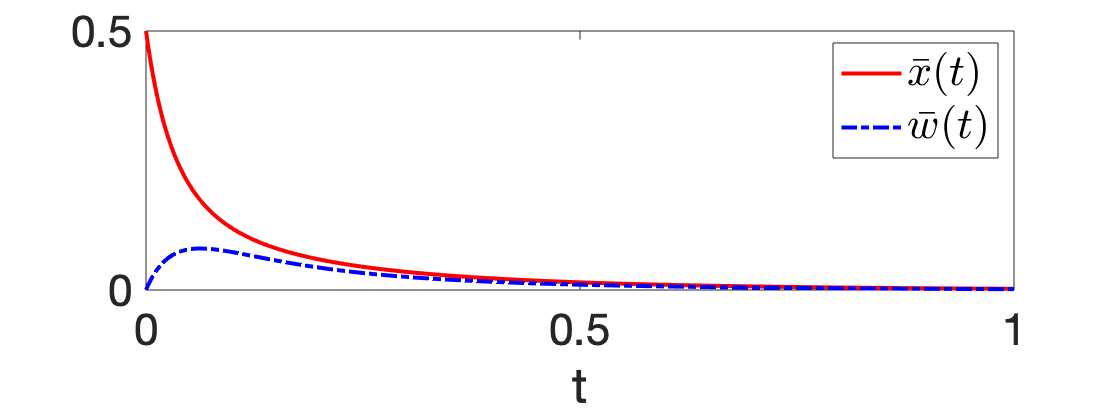}
      \caption{Eradicated}\label{fig:extinction}
    \end{subfigure}
    \begin{subfigure}[b]{.49\columnwidth}
    \includegraphics[width=\columnwidth]{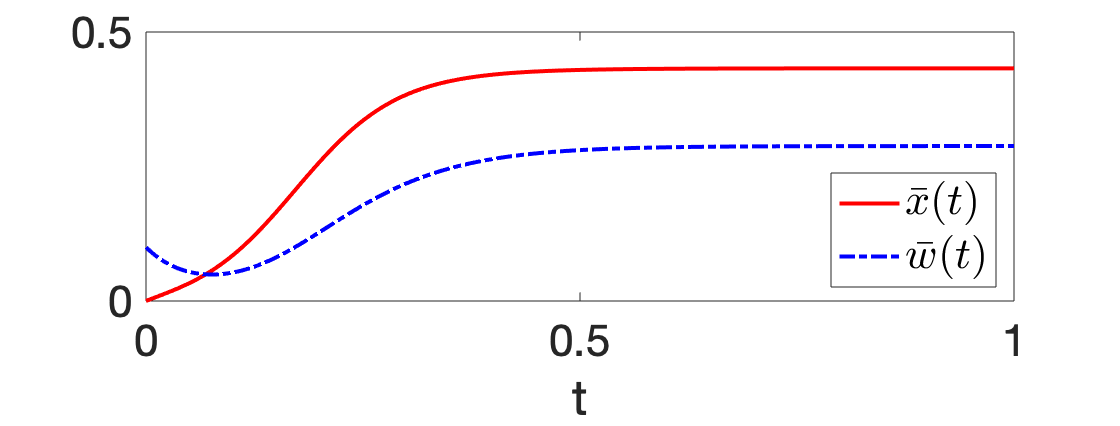}
      \caption{Endemic}\label{fig:endemic}
    \end{subfigure}
    \caption{Simulations with different outcomes. In (a) we have $\rho(D_f^{-1} B_f) < 1$, 
    eradicating the virus asymptotically. In (b) we have $\rho(D_f^{-1} B_f) > 1$, so the virus becomes endemic. 
    }
    \label{fig:sim}
\end{figure}

\begin{figure}[t!]
	\centering
    \begin{minipage}[c]{0.5\columnwidth}
        \includegraphics[width=\columnwidth]{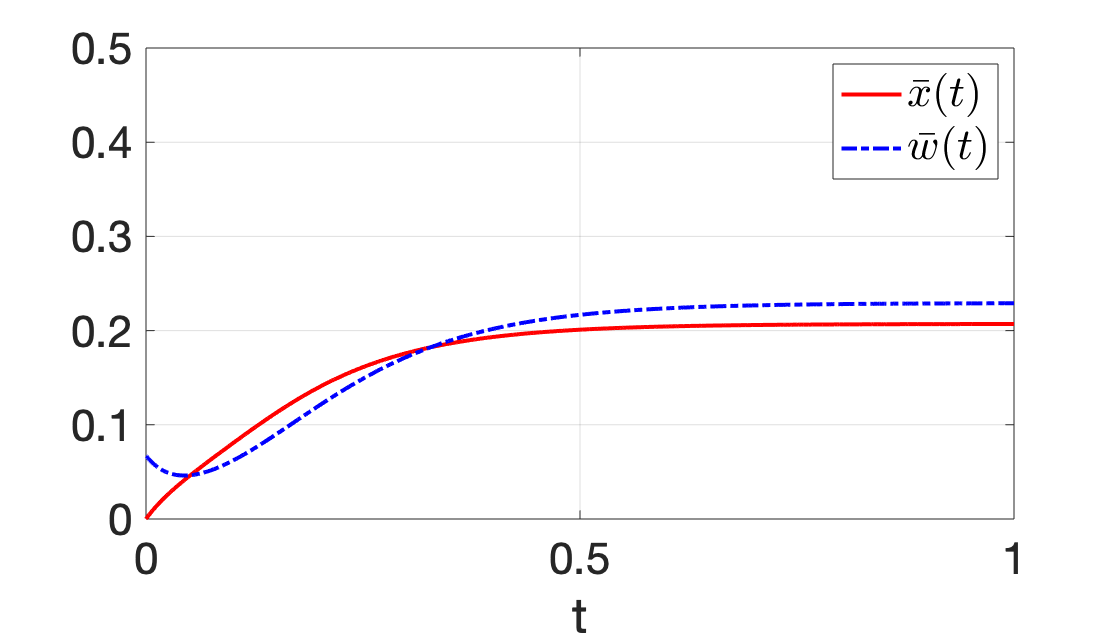}
    \end{minipage}\hfill
    \begin{minipage}[c]{0.5\columnwidth}
        \caption{Simulation employing Assumption~\ref{assum:sys:params}. Here $D_w$ has some zero entries, in violation of Assumption~\ref{assum:base}. Nonetheless, the simulation converges to some endemic equilibrium.} 
        \label{fig:assum1_endem}
    \end{minipage}
\end{figure}

For all simulations, we consider a network of $10$ population nodes and $15$ resource nodes. This network is depicted in Fig.~\ref{fig:network}, 
with population nodes as squares and resource nodes as circles. We denote the average infection proportion of the virus across the population nodes by $\Bar{x}(t)$, and the average contamination across the resource nodes by $\Bar{w}(t)$. The terms $\beta_{ij}$, $\beta^w_{ij}$, and $\alpha_{ij}$ are all binary, i.e. equal to one whenever nodes $i$ and $j$ are neighbors, for all simulations. For the simulations in Fig.~\ref{fig:sim} and Fig.~\ref{fig:assum1_endem} we set $c_{ij}^w$ to be binary 
which results in the network being irreducible.
By choosing
$D = 5I$, $D_w = 5I$, we see that $\rho(D_f^{-1} B_f) < 1$. 
Consequently, consistent 
with the result in Theorem~\ref{thm:GAS}, the virus is asymptotically eradicated across the network; see Fig.~\ref{fig:extinction}. 
Choosing $D = 2I$, $D_w = 2I$ results in $\rho(D_f^{-1}B_f) > 1$. Therefore, 
consistent with the result in Theorem~\ref{thm:equi}, the virus becomes endemic across all population and resource nodes, asymptotically approaching some positive equilibrium; see Fig.~\ref{fig:endemic}. Choosing $D = 4I$ and $D_w$ equal to a zero matrix, except for one diagonal entry equal to $100$, Assumption~\ref{assum:sys:params} is fulfilled but Assumption~\ref{assum:base} is violated. Therefore Theorem~\ref{thm:equi} does not apply, despite $\rho(D_f^{-1}B_f) > 1$, yet the virus still appears to converge to some positive equilibrium; see Fig.~\ref{fig:assum1_endem}.

\begin{figure}[t!]
  \centering
    \begin{subfigure}{.45\columnwidth}
    \includegraphics[width=\columnwidth]{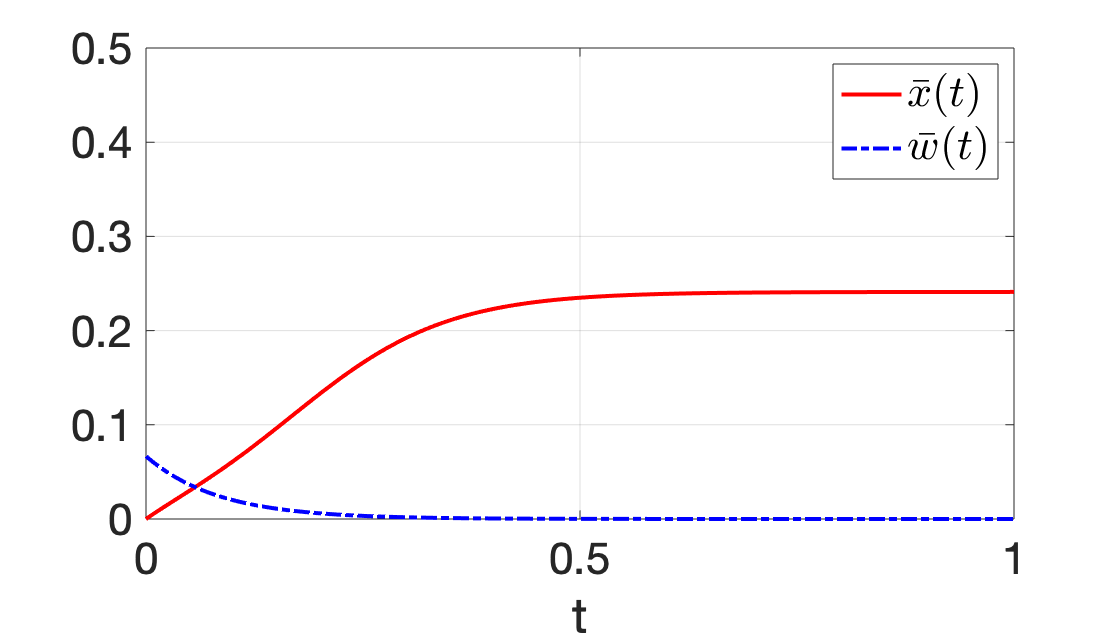}
    \subcaption{}
     \label{fig:nocontact}
    \end{subfigure}%
    \centering
    \begin{subfigure}{.45\columnwidth}
    \includegraphics[width=\columnwidth]{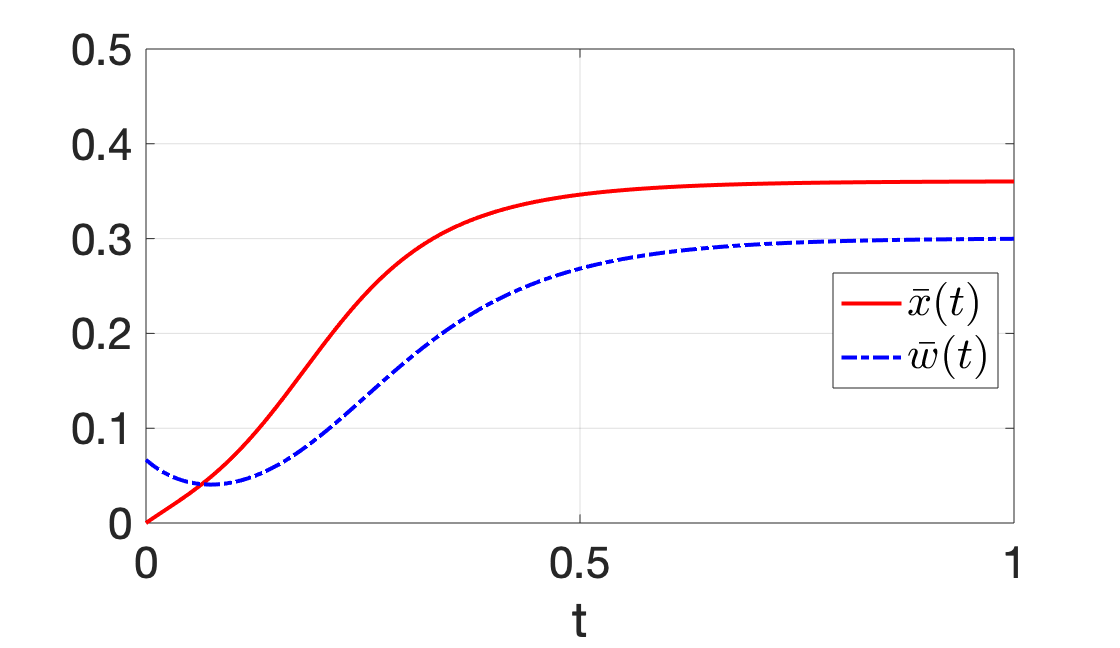}
    \subcaption{}
    \label{fig:contact}
    \end{subfigure}
    \caption{Simulations without and with population-to-resource contact. In (a) $C_w = 0$, so the contamination of the resource nodes decays to zero. In (b) $C_w$ is non-zero, and therefore the virus becomes endemic in the resource nodes as well.  
    }
    \label{fig:contact_nocontact}
   
\end{figure}

For the simulations depicted in Fig.~\ref{fig:contact_nocontact} we chose $D = 3I$, $D_w = 0.2I$.  
Since $D_w$ is a positive diagonal matrix, the resource network requires some non-zero $c_{ij}^w$ to sustain a positive level of contamination. 
Choosing $c_{ij}^w = 0$ for all $i, j$ ensures that the contamination across all resource nodes decays to zero; see the blue curve in Fig.~\ref{fig:nocontact}. However, $B$ is an irreducible matrix, and we still have $\rho(D^{-1}B) > 1$. Therefore, the infection levels in the population network converge to an endemic equilibrium, consistent 
with the results in~\cite{FallMMNP07,liu2019analysis}; 
see the red curve in Fig.~\ref{fig:nocontact}. 
Setting $c_{ij}^w$ to be binary as before 
results in the contamination of the resource network converging to a positive equilibrium; see the blue curve in Fig.~\ref{fig:contact}.
\seb{Consistent with the result in Proposition~\ref{prop:endemic_largerwithres}, it can be seen that in the absence of contamination in the resources, the endemic state in the population is smaller, whereas if the resources are also contaminated then the endemic state in the population is larger; see Fig.~\ref{fig:nocontact} and~\ref{fig:contact}, respectively}.

\section{Conclusion}\label{sec:conclusion}

In this paper, we have developed a multi-network-dependent, continuous-time SIWS epidemic model, also referred to as a layered networked SIWS model. This model captures a networked system, which can be interpreted as individual people or multiple groups of individuals, coupled with an infrastructure network, which can be understood as a contaminated water (or some other utility) distribution network. 
We have analyzed the stability of the healthy state, both locally and globally. We compared the basic reproduction number of the model with the standard networked SIS model without a pathogen. We have established conditions for the existence, uniqueness, and 
stability
of an endemic equilibrium. We have also provided a necessary and  sufficient condition 
for the healthy state to be the only equilibrium of this model. Lastly, we have established conditions under which the initial infection levels of the shared resources could be recovered based on the measurements of the infection levels of the population.

One line of future investigation could focus on understanding the spread of diseases in infrastructure networks with time-varying topologies. Another problem of interest would be to develop control algorithms that exploit the topology of the infrastructure network for  virus mitigation. Still on the topic of control of epidemics, it would be interesting to mitigate (resp. eradicate) epidemics subject to constraints on the availability of healing resources. 

\bibliographystyle{IEEEtran}
\bibliography{bib}

\section*{Appendix}
\subsection*{Proof of Prosition~\ref{jjj}}
To prove Proposition~\ref{jjj}, we need the following lemma. 

\begin{lem}
{\cite[Proposition~1]{liu2019analysis}}
Suppose that $N$ is an irreducible nonnegative matrix in $\R^{n\times n}$ and $\Lambda$ is a negative diagonal matrix in $\R^{n\times n}$.
Let $M=N + \Lambda$.
Then, $s(M)<0$ if and only if $\rho(-\Lambda^{-1}N)<1$,
$s(M)=0$ if and only if $\rho(-\Lambda^{-1}N)=1$,
and $s(M)>0$ if and only if $\rho(-\Lambda^{-1}N)>~1$.
\label{ff}
\end{lem}

{\em Proof of Proposition~\ref{jjj}:}
By Assumption~\ref{para}, $\delta^w_{j}+ \sum_{k}\alpha_{kj}>0$  for all $j\in[m]$, $D_f$ is invertible. 
From Lemma~\ref{ff}, the condition $\rho(D_f^{-1}B_f)<1$ is equivalent to $s(B_f-D_f)<0$, which implies that $J(\0,\0)$ is a continuous-time stable matrix. Thus, by Lyapunov's indirect method  the healthy state $(\0,\0)$ of \eqref{x}-\eqref{w} is locally exponentially stable.~$\square$

\subsection*{Proof of Theorem~\ref{thm:GAS}}
To prove the claim in Theorem~\ref{thm:GAS}, we need the following lemmas.

\begin{lem} 
{\cite[Lemma~2.3]{varga}}
Suppose that $M$ is an irreducible Metzler matrix. Then,
$s(M)$ is a simple eigenvalue of $M$ and there exists a unique (up to scalar multiple) vector $x\gg \0$ such that
$Mx=s(M)x$.
\label{metzler0}\end{lem}

\begin{lem}
{\cite[Proposition~2]{rantzer}}
Suppose that $M$ is an irreducible Metzler matrix such that $s(M)<0$.
Then, there exists a positive diagonal matrix $P$ such that
$M^\top P+PM \prec 0$.
\label{less}\end{lem}

\begin{lem}
{\cite[Lemma A.1]{khanafer2016stability}}
Suppose that $M$ is an irreducible Metzler matrix such that $s(M)=0$.
Then, there exists a positive diagonal matrix $P$ such that
$M^\top P+PM \preccurlyeq 0$.
\label{equal}\end{lem}

\begin{lem}{\cite{khalil}}
Let $\tilde x$ be an equilibrium of $\dot x(t)  = f(x(t))$ and $\scr{E}\subset \scr{D}$ be a bounded domain
containing $\tilde x$. Let $V:\scr{E}\rightarrow\R$ be a continuously differentiable function
such that $V(\tilde x)=0$, $V(x)>0$ in $\scr{E}\setminus \{\tilde x\}$, $\dot V(\tilde x)=0$,
and $\dot V(x)<0$ in $\scr{E}\setminus \{\tilde x\}$. If $\scr{E}$ is an invariant set,
then the equilibrium $\tilde x$ is asymptotically stable with the domain of attraction $\scr{E}$.
\label{lya}
\end{lem}

{\em Proof of Theorem \ref{thm:GAS}:}
Recalling the notation in \eqref{eq:z}, we first consider the case when $\rho(D_f^{-1}B_f)<1$.
By Lemma~\ref{ff}, in this case, $s(B_f-D_f)<0$.
Since $(B_f-D_f)$ is an irreducible Metzler matrix, by Lemma \ref{less},
there exists a positive diagonal matrix $P$ such that
$(B_f-D_f)^\top P+P(B_f-D_f)$ is negative definite.
Consider the Lyapunov function 
$V(z(t))=z(t)^\top Pz(t)$.
Then, from \eqref{x}-\eqref{eq:full}, 
when $z(t)\neq \0$, we have
\begin{align*}
\dot V(z(t)) &= 2z(t)^\top P\dot z(t) \\
&= 2z(t)^\top P(B_f-D_f)z(t) \\
& \;\;\;\; + 2z(t)^\top
P\left[\begin{array}{cc}
               -X(t)B & -X(t)B_w\\
               0 & 0 \end{array}\right]
z(t) \\
&<  -2z(t)^\top
P\left[\begin{array}{cc}
               X(t)B & X(t)B_w\\
               0 & 0 \end{array}\right]
z(t) \\
& \leq  0,
\end{align*}
where the strict inequality holds by Lemma \ref{less} 
since 
$2z(t)^\top P(B_f-D_f)z(t) = z(t)^\top (B_f-D_f)^\top P+P(B_f-D_f)z(t)$.
Thus, in this case, $\dot V(z(t))<0$ if $z(t)\neq \0$.
From Lemma~\ref{box} 
and Lemma~\ref{lya}, 
the healthy state
is asymptotically stable with 
domain of attraction $\mathcal D$, with $\mathcal D$ given
in~\eqref{eq:set:D}.


Next we consider the case when $\rho(D_f^{-1}B_f)=1$. By Lemma~\ref{ff}, $s(B_f-D_f)=0$.
Since $(B_f-D_f)$ is an irreducible Metzler matrix, by Lemma~\ref{equal},
there exists a positive diagonal matrix $Q$ such that
$(B_f-D_f)^\top Q+Q(B_f-D_f)$ is negative semi-definite.
Consider the Lyapunov function $V(z(t))=z(t)^\top Qz(t)$.
Then, from \eqref{x}-\eqref{eq:full},
we have
\begin{align*}
\dot V(z(t)) 
&= 2z(t)^\top Q(B_f-D_f)z(t) \\
& \;\;\;\; + 2z(t)^\top
Q\left[\begin{array}{cc}
               -X(t)B & -X(t)B_w\\
               0 & 0 \end{array}\right]
z(t) \\
&\le   -2z(t)^\top
Q\left[\begin{array}{cc}
               X(t)B & X(t)B_w\\
               0 & 0 \end{array}\right]
z(t) \\
&=   -2z(t)^\top
\left[\begin{array}{cc}
               Q_1 & 0\\
               0 & Q_2 \end{array}\right]
    \left[\begin{array}{cc}
               X(t)B & X(t)B_w\\
               0 & 0 \end{array}\right]
z(t) \\
&= -2\left(x(t)^\top Q_1 X(t) Bx(t) + x(t)^\top Q_1 X(t)B_w w(t) \right) \\
& \leq  0,
\end{align*}
where $Q_1$ is the $n$th principal subarray of $Q$, which is an $n\times n$ positive diagonal matrix, and $Q_2$ is the $m\times m$ positive diagonal matrix that is composed of the rest of the block diagonal entries of $Q$. 
We claim that $\dot V(z(t))<0$ if $z(t)\neq \0$.
To establish this claim, we first consider the case when $z(t)\gg \0$. 
Since $B_f$ is irreducible and non-negative we have $B_f z(t) \gg \0$. As such, $B x(t) + B_w w(t) \gg \0$, and due to $Q_1$ being a positive diagonal matrix, it follows that $x(t)^\top Q_1 X(t) (B x(t) + B_w w(t)) > 0$. Thus, $\dot V(z(t))<0$.

Next we consider the case when $z(t)>\0$ and $z(t)$ has at least one zero entry.
If $(B_f-D_f)^\top Q+Q(B_f-D_f)$ does not have an eigenvalue at zero, then
$(B_f-D_f)^\top Q+Q(B_f-D_f)$ is negative definite, which implies that
$z(t)^\top \left((B_f-D_f)^\top Q+Q(B_f-D_f)\right)z(t)<0$ when
$z(t)>\0$ and, thus, in this case,
\begin{align*}
\dot V(z(t)) 
&= 2z(t)^\top Q(B_f-D_f)z(t) \\
& \;\;\;\; + 2z(t)^\top
Q\left[\begin{array}{cc}
               -X(t)B & -X(t)B_w\\
               0 & 0 \end{array}\right]
z(t) \\
&\le   2z(t)^\top Q(B_f-D_f)z(t)  <  0.
\end{align*}
Now suppose that $(B_f-D_f)^\top Q+Q(B_f-D_f)$ has an eigenvalue at zero.
Since $(B_f-D_f)$ is an irreducible Metzler matrix and $Q$ is a positive diagonal matrix,
$(B_f-D_f)^\top Q+Q(B_f-D_f)$ is a symmetric irreducible Metzler matrix.
Since $(B_f-D_f)^\top Q+Q(B_f-D_f)$ is negative semi-definite,
$s((B_f-D_f)^\top Q+Q(B_f-D_f))=0$. By Lemma~\ref{metzler0},  zero is a simple eigenvalue of
$(B_f-D_f)^\top Q+Q(B_f-D_f)$ and it has a unique (up to scalar multiple) strictly positive eigenvector
corresponding to the eigenvalue zero. Thus, $z(t)^\top \left((B_f-D_f)^\top Q+Q(B_f-D_f)\right)z(t)<0$ when
$z(t)>\0$ and $z(t)$ has at least one zero entry (because the only vector for which it equals zero is the strictly positive eigenvector).
Therefore, $\dot V(z(t))<0$ if $z(t)\neq \0$.
From 
Lemma \ref{box} 
and Lemma \ref{lya},
the healthy state
is asymptotically stable with 
domain of attraction 
$\mathcal{D}$, with $\mathcal D$
given in~\eqref{eq:set:D}.~$\square$ 

\subsection*{Proof of Theorem~\ref{thm:equi}}
To prove the claim in Theorem~\ref{thm:equi}, 
%
we will be making use of the following variants of the Perron-Frobenius theorem for irreducible matrices.
\begin{lem} \label{lem:perron_frob}
\cite[Chapter 8.3]{meyer2000matrix} \cite[Theorem~2.7]{varga}
Suppose that $N$ is an irreducible nonnegative matrix. Then,
\begin{enumerate}[label=(\roman*)]
    \item $r = \rho(N)$ is a simple eigenvalue of $N$. \label{item:perfrob_simpleeig}
    \item There is an eigenvector $\zeta \gg \textbf{0}$ corresponding to the eigenvalue $r$. \label{item:perfrob_pos_exists}
    \item $x > \textbf{0}$ is an eigenvector only if $Nx = rx$ and $x \gg \textbf{0}$. 
\end{enumerate}
\end{lem}
{\em Proof of Theorem \ref{thm:equi}:}
\noindent The proof is split in three parts: First we show existence of an endemic equilibrium provided the conditions in Theorem~\ref{thm:equi} are satisfied. Subsequently, we show that this equilibrium is unique, and that for all non-zero initial conditions the dynamics converge asymptotically to this equilibrium.\\ 

\noindent \textit{Part 1 -Proof of existence}\\Note that if $z \geq \textbf{0}$, $\diag(D_f^{-1} B_f z)$ is a nonnegative diagonal matrix, and therefore the inverse of $(I + \diag(D_f^{-1} B_f z))$ exists. Define a map $T(z): \mathbb{R}_+^{n+m} \rightarrow \mathbb{R}_+^{n+m}$ such that
\begin{equation*}
T(z) = (I + \diag(D_f^{-1} B_f z))^{-1} (D_f^{-1} B_f z + \diag(D_f^{-1} B_f z) 
\begin{bmatrix}
\textbf{0} \\
w
\end{bmatrix}
).
\end{equation*}
\noindent Observe that the components of $T(y)$ are
\begin{align*}
T_{i}(z) = &\frac{(D_f^{-1} B_f z)_{i}}{1 + (D_f^{-1} B_f z)_{i}}, \text{ for } i \in [n], \\
T_{j}(z) = &\frac{(D_f^{-1} B_f z)_{j} z_{j} + (D_f^{-1} B_f z)_{j}}{1 + (D_f^{-1} B_f z)_{j}}, \text{ for } j \in [n+m] \backslash [n].
\end{align*}
\noindent Note that the scalar function $s/(1+s)$ is increasing in $s$, and that $D_f^{-1} B_f$ is a nonnegative matrix. Therefore, $v \geq z$ implies $T(v) \geq T(z)$. Notice that a fixed point of $T(z)$ fulfills
\begin{equation} \label{eq:fixed}
z = (I + \diag(D_f^{-1} B_f z))^{-1} (D_f^{-1} B_f z + \diag(D_f^{-1} B_f z)
\begin{bmatrix}
\textbf{0} \\
w
\end{bmatrix}
).
\end{equation}
\noindent Multiplying~\eqref{eq:fixed} by $(I + \diag(D_f^{-1} B_f z))$ gives us
\begin{equation} \label{eq:diagflip}
D_f^{-1} B_f z + \diag(D_f^{-1} B_f z) 
\begin{bmatrix}
\textbf{0} \\
w
\end{bmatrix} 
= (I + \diag(D_f^{-1} B_f z)) z. 
\end{equation}
\noindent Using the identity $\diag(u) v = \diag(v) u$,~\eqref{eq:diagflip} is equivalent to
\begin{equation} \label{eq:z_diag_add}
D_f^{-1} B_f z + \diag(
\begin{bmatrix}
\textbf{0} \\
w
\end{bmatrix}
) D_f^{-1} B_f z = (I + \diag(z) D_f^{-1} B_f) z.
\end{equation}
\noindent Recall that the definition of $X(z)$ means that subtracting
$\diag(
\begin{bmatrix}
\textbf{0} \\
w
\end{bmatrix}
) D_f^{-1} B_f z$
\noindent from~\eqref{eq:z_diag_add} yields
\begin{equation} \label{eq:X_D_commute}
 D_f^{-1} B_f z = (I + X(z) D_f^{-1} B_f) z.
\end{equation}
\noindent Since $X(z)$ and $D_f^{-1}$ are diagonal matrices, they commute. Furthermore, by pre-multiplying~\eqref{eq:X_D_commute} with $D_f$, and suitably rearranging terms, we obtain
\begin{equation} \label{eq:equi_equa}
 ( - D_f + (I - X(z)) B_f) z = 0. 
\end{equation}
\noindent A solution of equation~\eqref{eq:equi_equa} is clearly an equilibrium of~\eqref{eq:full}. As such, it suffices to show that $T(z)$ has a fixed point $\Tilde{z} \gg \textbf{0}$. We will now show that at least one such fixed point exists.


We have $\rho(D_f^{-1} B_f) > 1$. Note that $D_f^{-1} B_f$ is an irreducible nonnegative matrix. Hence, by 
Lemma~\ref{metzler0},
$\lambda^* = \rho(D_f^{-1} B_f)$ is a simple eigenvalue of $D_f^{-1} B_f$ 
and
the eigenspace of $\lambda^*$ is spanned by a vector $y^* \gg \textbf{0}$. Then, since $\lambda^* > 1$, there exists some $\epsilon > 0$ such that, for all $i \in [n+m]$, we have $\epsilon z_i^* \leq (\lambda^* - 1)/\lambda^*$, which implies that $1 \leq \lambda^*/(1 + \lambda^* \epsilon z_i^*)$. Hence, $\epsilon z_i^* \leq \lambda^* \epsilon z_i^* / (1 + \lambda^* \epsilon z_i^*)$, and thus
\begin{equation} \label{eq:i_comp}
 \epsilon z_i^* \leq \frac{(D_f^{-1} B_f \epsilon z^*)_i}{1 + (D_f^{-1} B_f \epsilon z^*)_i}, \text{ for all } i \in [n].
\end{equation}
\noindent Noting that $(D_f^{-1} B_f \epsilon z^*)_{j} \epsilon z_{j}^* > 0$ for all $j \in [n+m] \backslash [n]$, we also have
\begin{equation} \label{eq:n+m_comp}
\epsilon z_{j}^* \leq \frac{(D_f^{-1} B_f \epsilon z^*)_{j} \epsilon z_{j}^* + (D_f^{-1} B_f \epsilon z^*)_{j}}{1 + (D_f^{-1} B_f \epsilon z^*)_{j}}, 
\end{equation}
for all $j \in [n+m] \backslash [n]$.
\noindent Due to the inequalities~\eqref{eq:i_comp} and~\eqref{eq:n+m_comp}, we have $T(\epsilon z^*) \geq \epsilon z^*$. Since $z \geq r$ implies 
$T(z) \geq T(r)$, it follows that for any $z \geq \epsilon z^*$ we have $T(z) \geq \epsilon z^*$. 
Define the vector
\begin{gather*} 
 \textbf{z}:=
 \begin{bmatrix}
 \textbf{1} \\  \textbf{w} 
 \end{bmatrix},
\end{gather*}
where $\textbf{w} := -(A_w - D_w)^{-1} C_w \textbf{1}$. Note that $(A_w - D_w)$ is invertible because of it being diagonally dominant. 
\noindent Consider $T_{i}(\textbf{z})$ for $i \in [n]$ while noting that $s/(1+s)$ is bounded from above by $1$ for any positive $s$. Then
\begin{equation} \label{eq:i_one_ineq}
T_{i}(\textbf{z}) = \frac{(D_f^{-1} B_f \textbf{z})_{i}}{1 + (D_f^{-1} B_f \textbf{z})_{i}} \leq 1, \text{ for all } i \in [n].
\end{equation}
\noindent Before considering $T_{j}(\textbf{z})$ for $j \in [n+m] \backslash [n]$, first note that
\begin{align*}
    &
    (D_w - \diag (A_w))^{-1}
    \begin{bmatrix}
    C_w & A_w - \diag (A_w)
    \end{bmatrix}
    \,
    \begin{bmatrix}
    \textbf{1} \\
    \textbf{w}
    \end{bmatrix}
    \,
    \\
    &=  (D_w - \diag (A_w))^{-1} C_w \textbf{1} - (D_w - \diag (A_w))^{-1}\\
    &~~~~~~~~~~~~~~~~~~~~~~\times(A_w - \diag (A_w))  (A_w - D_w)^{-1} C_w \textbf{1} \\
    &= (D_w - \diag (A_w))^{-1} (I -(A_w - \diag (A_w)) (A_w - D_w)^{-1}) \\
    &~~~~~~~~~~~~~~~~~~~~~~\times C_w \textbf{1} \\
    &= (D_w - \diag (A_w))^{-1} ((A_w - D_w) - (A_w - \diag (A_w))) \\ 
    &~~~~~~~~~~~~~~~~~~~~~~~~~~~~~~~~~~~~~~~~~~~~~~\times(A_w - D_w)^{-1} C_w \textbf{1} \\
    &= -(D_w - \diag (A_w))^{-1} (D_w - \diag (A_w)) (A_w - D_w)^{-1} C_w \textbf{1} \\
    &= -(A_w - D_w)^{-1} C_w \textbf{1} \\
    &= \textbf{w}
\end{align*}
\noindent Hence,
\begin{equation} \label{eq:n+m_one_ineq}
T_{j}(\textbf{z}) = \frac{\textbf{z}_j(1 + \textbf{z}_j)}{1 + \textbf{z}_j} = \textbf{z}_j, \text{ for all } j \in [n+m] \backslash [n].
\end{equation}
\noindent Due to~\eqref{eq:i_one_ineq} and~\eqref{eq:n+m_one_ineq}, we have $T(\textbf{z}) \leq \textbf{z}$. Since $v \geq w$ implies $T(v) \geq T(w)$, it follows that $T(z) \leq \textbf{z}$ if $z \leq \textbf{z}$. By Brouwer's fixed-point theorem, there is at least one fixed point of $T(z)$ in the domain $\{ z : \epsilon z^* \leq z \leq \textbf{z}\}$. In conclusion, the map $T(z)$ has at least one fixed point in the domain $\{ z : \epsilon z^* \leq z \ll \textbf{z}\}$, and therefore 
\eqref{eq:full} has at least one equilibrium $\Tilde{z} \gg \textbf{0}$.~$\square$ \\

\noindent \textit{Part~2 -- Proof of uniqueness}

\noindent We will now prove that the 
endemic equilibrium is unique. Suppose that there are two 
endemic equilibria, $\Tilde{z}$ and $\Tilde{\mathbf{z}}$. 
Note that, by similar arguments as in \cite[Lemma~6]{axel2020TAC},  $\Tilde{z} \gg \textbf{0}$ and $\Tilde{\mathbf{z}} \gg \textbf{0}$. 
Let $\kappa = \max_{i \in [n+m]} \Tilde{z}_i / \Tilde{\mathbf{z}}_i$. It turns out that $\kappa$ is given by
\vspace{-1ex}
\begin{equation} \label{eq:unique_single-virus_kappadef}
\kappa = \max_{i \in [n]} \Tilde{z}_i/\Tilde{\mathbf{z}}_i.
\end{equation}

\vspace{-1.5ex}

\noindent
To see this, assume by way of contradiction that $\kappa = \Tilde{z}_{n+j}/\Tilde{\mathbf{z}}_{n+j}$ for some $j \in [m]$, and 
thus 
$\kappa > \Tilde{z}_i/\Tilde{\mathbf{z}}_i$, for all $i \in [n]$. Since both $\Tilde{z}$ and $\Tilde{\mathbf{z}}$ are equilibria of system~\eqref{eq:full}, it follows that, for each $j \in [m]$
\vspace{-1.5ex}
\begin{equation} \label{eq:shared_resource_weighted_average}
\begin{split}
\Tilde{z}_{n+j} &=  \textstyle \sum_i^{n} c_i \Tilde{x}_i + \textstyle \sum_{k, k\neq j}^{m} \alpha_k \Tilde{w}_k,\\
\Tilde{\mathbf{z}}_{n+j} &= \textstyle \sum_i^{n} c_i \Tilde{\mathbf{x}}_i  + \textstyle \sum_{k, k\neq j}^{m} \alpha_k \Tilde{\mathbf{w}}_k.
\end{split}
\end{equation}

\vspace{-1.5ex}

\noindent
Since we 
have
that $\kappa > \Tilde{z}_i/\Tilde{\mathbf{z}}_i$, for all $i \in [n]$, then 
$\kappa \Tilde{\mathbf{z}}_i > \Tilde{z}_i$, for all $i \in [n]$. Since 
by assumption
$\kappa = \Tilde{z}_{n+j}/\Tilde{\mathbf{z}}_{n+j}$ for some $j \in [m]$, it follows that, for each $k \in [m]$, $\Tilde{z}_{n+k} \leq \kappa \Tilde{\mathbf{z}}_{n+k}$.  Then,~\eqref{eq:shared_resource_weighted_average} yields
\vspace{-1ex}
\begin{align*}
\Tilde{z}_{n+j} &= \textstyle \sum_i^{n} c_i \Tilde{x}_i + \textstyle \sum_{k, k\neq j}^{m} \alpha_k \Tilde{w}_k \nonumber \\
&< \textstyle \kappa \sum_i^{n} c_i \Tilde{\mathbf{x}}_i  + \textstyle \sum_{k, k\neq j}^{m} \alpha_k \Tilde{\mathbf{w}}_k \nonumber \\ 
&= \kappa \Tilde{\mathbf{z}}_{n+j}. 
\end{align*}

\vspace{-1.5ex}

\noindent
Hence, for all $j \in [m]$, $\kappa > \Tilde{z}_{n+j}/\Tilde{\mathbf{z}}_{n+j}$, which contradicts the assumption that $\kappa = \Tilde{z}_{n+j}/\Tilde{\mathbf{z}}_{n+j}$, for some $j \in [m]$. 
Therefore, $\kappa$ must be given by 
\eqref{eq:unique_single-virus_kappadef}.
Now, by~\eqref{eq:unique_single-virus_kappadef} we know that $\Tilde{z} \leq \kappa \Tilde{\mathbf{z}}$. For some $j \in [n]$ we have $\Tilde{z}_j = \kappa \Tilde{\mathbf{z}}_j$. 
Assume, by way of contradiction, that $\kappa > 1$. Then, since an equilibrium of~\eqref{eq:full} also constitutes a fixed point of $T(z)$, we have
%
\vspace{-1ex}
\begin{align}
\Tilde{z}_j &= (D_f^{-1} B_f \Tilde{z})_{j} / (1 + (D_f^{-1} B_f \Tilde{z})_{j}) \nonumber \\
&\leq (D_f^{-1} B_f \kappa \Tilde{\mathbf{z}})_{j} / (1 + (D_f^{-1} B_f \kappa \Tilde{\mathbf{z}})_{j}) \label{eq:singleequi_unique_kappaybigger} \\
&< \kappa (D_f^{-1} B_f \Tilde{\mathbf{z}})_{j} / (1 + (D_f^{-1} B_w \Tilde{\mathbf{z}})_{j}) \label{eq:singleequi_unique_kappabiggerthanone} \\
&= \kappa \Tilde{\mathbf{z}}_j \label{eq:singleequi_unique_alsoequilibrium}  \\
&= \Tilde{z}_j, \label{eq:singleequi_unique_j_hasequality}
\end{align}

\vspace{-1.5ex}

\noindent
where~\eqref{eq:singleequi_unique_kappaybigger} follows from $\Tilde{z} \leq \kappa \Tilde{\mathbf{z}}$ and that $T(v) \geq T(w)$ whenever $v \geq w$,~\eqref{eq:singleequi_unique_kappabiggerthanone} follows from the assumption $\kappa > 1$, and~\eqref{eq:singleequi_unique_alsoequilibrium} follows from the fact that $\Tilde{\mathbf{z}}$ is an equilibrium of~\eqref{eq:full}. Note that~\eqref{eq:singleequi_unique_j_hasequality} is a contradiction, following from our assumption that $\kappa >1$. Hence, $\kappa \leq 1$, meaning that $
\Tilde{z}\leq \Tilde{\mathbf{z}}$. Switching the roles of $\Tilde{z}$ and $\Tilde{\mathbf{z}}$, we see that $\Tilde{\mathbf{z}} \leq \Tilde{z}$. Therefore, $\Tilde{z} = \Tilde{\mathbf{z}}$, and thus the equilibrium is unique.\\

\noindent \textit{Part 3: Proof of asymptotic convergence}\\
Let $\Delta z(t) = z(t) - \tilde{z}$, for $z(t) \in \mathcal D$. Also let, $X(\Delta z(t)) = X(z(t)) - X(\tilde z)$. Therefore, it follows that:
\vspace{-1ex}
\begin{align}
\Delta \dot{z}(t) &= (- D_f + (I - X(\Delta z) - X(\Tilde{z}))B_f) (\Delta z + \Tilde{z}) \nonumber \\
&= (- D_f + (I - X(\Tilde{z}))B_f) \Delta z - X(\Delta z)B_f z \label{eq:use_equi_is_zero} 
\\
&= (- D_f + (I - X(\Tilde{z}))B_f - X(B_f z)) \Delta z, \label{eq:use_X_diag_relation}
\end{align}

\vspace{-1.5ex}

\noindent where~\eqref{eq:use_equi_is_zero} is obtained by noting that since $\tilde z$ is the unique endemic equilibrium of system~\eqref{eq:full}  in $\mathcal D$, 
\begin{align}
(- D_f + (I - X(\Tilde{z}))B_f) \Tilde{z} = \textbf{0} \label{eq:equi_necess_single-virus}
\end{align}
Equation~\eqref{eq:use_X_diag_relation} follows by noting that, for any two vectors $u$ and $v$, $X(u)v = X(v)u$.\\
By Assumption~\ref{assum:base}, it is immediate that $D$ and $D_w$ are positive diagonal matrices. Since $\diag{(A_w)}= -\sum_k \alpha_{kj}$, it follows that  $-\diag{(A_w)}= \sum_k \alpha_{kj}$. Hence Assumption~\ref{assum:base} also ensures that  $-\diag{(A_w)}$ is nonnegative, and as a consequence, $D_w -\diag{(A_w)}$ is a positive diagonal matrix. Therefore, $D_f$ is a positive diagonal matrix, and, hence, is invertible. Consequently,~\eqref{eq:equi_necess_single-virus} can be rewritten as:
\begin{equation}\label{eq:equi_proper_single-virus}
    (I - X(\Tilde{z})) D_f^{-1} B_f \Tilde{z} = \Tilde{z}.
\end{equation}
Observe that, since
$\tilde{z} \ll \mathbf{1}$, $I - X(\Tilde{z})$ is a positive diagonal matrix.  Since, by assumption, $B_f$ is irreducible, and by Assumption~\ref{assum:base} $B_f$ is nonnegative, it follows that $(I - X(\Tilde{z})) D_f^{-1} B_f$  is nonnegative irreducible. Since $\tilde{z} \gg \mathbf{0}$,  item iii) in Lemma~\ref{lem:perron_frob} yields $\rho((I - X(\Tilde{z})) D_f^{-1} B_f) =1$, which, from Lemma~\ref{ff}, further implies that $s(-D_f +(I - X(\Tilde{z}))B_f) =0$. Therefore, since the matrix $(-D_f +(I - X(\Tilde{z}))B_f)$ is irreducible Metzler, by Lemma~\ref{equal} there exists a positive diagonal matrix $Q$ such that  $\Psi \coloneqq (- D_f + (I - X(\Tilde{z}))B_f)^T Q + Q (- D_f + (I - X(\Tilde{z}))B_f) \preccurlyeq~0$.\\
Define the Lyapunov function candidate $V(\Delta z(t)) = \Delta z(t)^T Q \Delta z(t)$, with $z(t) \in \mathcal{D} \setminus \{ \textbf{0}\}$ as the domain. Since $Q$ is a positive diagonal matrix, $V(\Delta z(t)) \succ 0$.
Differentiating $V(\Delta z(t))$ with respect to $t$ yields:
\vspace{-1ex}
\begin{equation}
\dot{V}(\Delta z(t)) 
= \Delta z^T \Psi \Delta z - 2  \Delta z^T Q X(B_f z) \Delta z \label{eq:lyapunov_diff_singlevirusendemic}
\end{equation}

\vspace{-1ex}
\noindent where~\eqref{eq:lyapunov_diff_singlevirusendemic} makes use of~\eqref{eq:use_X_diag_relation}. The rest of the proof consists of showing that $\dot{V}(\Delta z(t)) < 0$ for all $z(t) \in \mathcal{D} \setminus \{ \textbf{0}\}$ such that $\Delta z(t) = z(t) - \Tilde{z} \neq \textbf{0}$. First, consider all $z(t) \in \mathcal{D} \setminus \{ \textbf{0}\}$ such that $z(t) \gg \textbf{0}$. Since $\Psi \preccurlyeq 0$, it is immediate that 
\vspace{-1ex}
\begin{align}
\dot{V}(\Delta z(t)) &\leq - 2 \Delta z(t)^T Q X(B_f z(t)) \Delta z(t). \label{eq:using_negative_semidefiniteness}
\end{align}

\vspace{-1.5ex}

\noindent
Since $z(t) \gg \textbf{0}$, $Q X(B_f z(t))$ is a positive diagonal matrix, and thus $Q X(B_w y(t)) \succ 0$. Therefore, from~\eqref{eq:using_negative_semidefiniteness} it is clear that $\dot{V}(\Delta z(t)) < 0$ for all $z(t) \in \mathcal{D} \setminus \{ \textbf{0}\}$ such that $z(t) \gg \textbf{0}$, $\Delta z(t) = z(t) - \Tilde{z} \neq \textbf{0}$. Now, consider all $z(t) \in \mathcal{D} \setminus \{ \textbf{0}\}$ such that $z(t) > \textbf{0}$, $z_i(t) = 0$ for some $i \in [n+m]$. Since $z(t) > \textbf{0}$, $Q X(B_w y(t))$ is a nonnegative diagonal matrix. Then~\eqref{eq:lyapunov_diff_singlevirusendemic} can be bounded by
\vspace{-1ex}
\begin{align}
\dot{V}(\Delta z(t)) &\leq \Delta z(t)^T \Psi \Delta z(t). \label{eq:using_nonparallelism}
\end{align}

\vspace{-1.5ex}

\noindent
Since $(- D_f + (I - X(\Tilde{z}))B_f)$ is an irreducible Metzler matrix and $Q$ is a positive diagonal matrix, $\Psi$ is an irreducible Metzler matrix. 
Employing~\eqref{eq:equi_necess_single-virus}, we see that
\vspace{-1ex}
\begin{equation} \label{eq:cool_trick}
\Tilde{z}^T \Psi \Tilde{z} = 0. 
\end{equation}

\vspace{-1.5ex}

\noindent
Lemma~\ref{metzler0} 
stipulates that $r \coloneqq s(\Psi)$ is a simple eigenvalue of $\Psi$. Due to $\Psi \preccurlyeq 0$ and the Rayleigh-Ritz Theorem \cite[Theorem~4.2.2]{horn2012matrix}, it follows from~\eqref{eq:cool_trick} 
that $r=0$, and that $\Tilde{z}$ spans the eigenspace of $r$. Hence, due to $\Tilde{z} \gg \textbf{0}$, and $z(t) > \textbf{0}$ with $z_i(t) = 0$ for some $i \in [n+m]$, $z(t)$ can not be parallel to $\Tilde{z}$. Consequently, $\Delta z(t)$ can not be parallel to $\Tilde{z}$. By the Rayleigh-Ritz Theorem \cite[Theorem~4.2.2]{horn2012matrix}, $x^T \Psi x = r x^T x$ only if $x$ is parallel to $\Tilde{z}$, and $x^T \Psi x < r x^T x$ otherwise. Therefore, $r = 0$ together with~\eqref{eq:using_nonparallelism} gives us $\dot{V}(\Delta z(t)) < 0$ for all $z(t) \in \mathcal{D} \setminus \{ \textbf{0}\}$ such that $z(t) \gg \textbf{0}$ and $z(t) - \Tilde{z} \neq \textbf{0}$. 

Thus, we have $\dot{V}(\Delta z(t)) < 0$ for all $z(t) \in \mathcal{D} \setminus \{ \textbf{0}\}$ such that $\Delta z(t) = z(t) - \Tilde{z} \neq \textbf{0}$, and it is clear that $\dot{V}(\textbf{0}) = 0$. Therefore, $\dot{V}(\Delta y(t)) \prec 0$ for $y(t) \in \mathcal{D} \setminus \{ \textbf{0}\}$. Finally, for reasons similar to that in \cite[Lemma~7]{axel2020TAC}, we have that $\mathcal{D} \setminus \{ \textbf{0}\}$ is a positively invariant set with respect to~\eqref{eq:full}. Thus, we see that $V(\Delta z(t))$ meets the conditions for 
\cite[Theorem~4.1]{khalil2002nonlinear}
with respect to the shifted coordinates $\Delta z(t) = z(t) - \Tilde{z}$, for all $z(t) \in \mathcal{D} \setminus \{ \textbf{0}\}$. Hence, the unique endemic equilibrium $\Tilde{z}$ is asymptotically stable, with domain of attraction containing~$\mathcal{D} \setminus \{ \textbf{0}\}$.~$\square$ 

\vspace{-1ex}
\subsection*{Proof of Proposition~\ref{prop:endemic_largerwithres}}
In order to prove Proposition~\ref{prop:endemic_largerwithres}, we need the following result.
\begin{lem} \label{lem:axelTAC}
Consider system~\eqref{x_SIS} under Assumption~\ref{assum:noshared}. If $\rho(D^{-1}B)>1$, then there exists a unique endemic equilibrium $\tilde x$ such that $\textbf{0} \ll \tilde x \ll \textbf{1}$.
\end{lem}
\textit{Proof:} The result follows by particularizing  \cite[Theorem~3]{axel2020TAC} for the networked SIS model.~$\square$\\

\textit{Proof of Proposition~\ref{prop:endemic_largerwithres}:} By assumption, the matrices $B_f$ and $B$ are irreducible. Moreover, $\rho(D_f^{-1}B_f)>1$, and $\rho(D^{-1}B)>1$. Therefore, from Theorem~\ref{thm:equi}, and from Lemma~\ref{lem:axelTAC}, we know that there exists a unique endemic equilibrium $\hat z = [\begin{smallmatrix}\hat x \\ \hat w \end{smallmatrix}]$ for~\eqref{eq:full}, and  a unique endemic equilibrium $\tilde{x}$ for~\eqref{x_SIS}, respectively. 
Moreover, $\textbf{0} \ll \tilde{x} \ll \textbf{1}$.\\
\noindent Note that since $\tilde{x}$ is an equilibrium of \eqref{x_SIS}, 
we have
\begin{equation} \label{eq:equi_nores}
    (I-\tilde{X})D^{-1}B \tilde{x} = \tilde{x}.
\end{equation}
Consider a solution $z(t) = (x(t), w(t))$ to \eqref{eq:full} for $t \geq 0$, with $x_i(0) \in [0, 1]$ and $w_j(0) \geq 0$ for all $i \in [n]$, $j \in [m]$. By Lemma~\ref{lem:z_invariantset} we have $x_i(t) \in [0, 1]$ and $w_j(t) \geq 0$ for all $i \in [n]$, $j \in [m]$ and $t \geq 0$. Suppose that, for some $t \geq 0$, $x(t) \geq \tilde{x}$, with $x_i(t) = \tilde{x}_i$ for some $i \in [n]$. Then
\begin{align}
    \dot{x}_i (t) & = (1 - x_i(t))(D_f^{-1}B_f z(t))_i - x_i(t) \nonumber\\
    & = (1 - \tilde{x}_i)(D^{-1}B x(t) + D^{-1}B_w w(t))_i - \tilde{x}_i \nonumber\\
    & \geq (1 - \tilde{x}_i)(D^{-1}B \tilde{x})_i - \tilde{x}_i \label{eq:DinvBw_w_nonnegative_1} \\
    & = 0, \label{eq:equi_nores_zero}
\end{align}
where~\eqref{eq:DinvBw_w_nonnegative_1} follows from $D^{-1}B_w w(t) \geq \mathbf{0}$, and~\eqref{eq:equi_nores_zero} follows from~\eqref{eq:equi_nores}. 
Since the same argument holds for any $t$ and $i \in [n]$ we have $x(t) \geq \tilde{x}$ for all $t \geq 0$ if $x(0) \geq \tilde{x}$. 
Furthermore, due to (a):~$\tilde{x} \ll \mathbf{1}$, 
and (b):~$\hat{z}$ being the unique equilibrium of~\eqref{eq:full} with a region of attraction including $\{z = (x,w) : \mathbf{1} \geq x \geq \tilde{x}\}$, we must have $\hat{x} \geq \tilde{x}$. In order to show $\hat{x} \neq \tilde{x}$, assume by way of contradiction that $\hat{x} = \tilde{x}$. Note that
\begin{equation} \label{eq:hat_x_equi}
    \hat{x} = (I - \hat{X})(D^{-1}B \hat{x} + D^{-1}B_w \hat{w}).
\end{equation}
With the assumption that $\hat{x} = \tilde{x}$,~\eqref{eq:hat_x_equi} is equivalent to
\begin{align}
    \tilde{x} &= (I - \tilde{X})(D^{-1}B \tilde{x} + D^{-1}B_w \hat{w}) \nonumber \\
    & < (I - \tilde{X})D^{-1}B \tilde{x} \label{eq:DinvBw_w_nonnegative_2} \\
    & = \tilde{x}, \label{eq:using_equi_nores}
\end{align}
where \eqref{eq:DinvBw_w_nonnegative_2} is due to the following: (i):~$D_f^{-1}B_f$ is an irreducible Metzler matrix, (ii): $\hat{w} \gg \mathbf{0}$, and (iii):~$\tilde{x} \ll \mathbf{1}$, so we have $(I - \tilde{X})D^{-1}B_w \hat{w} > \mathbf{0}$. Moreover,~\eqref{eq:using_equi_nores} follows from~\eqref{eq:equi_nores}. Clearly,~\eqref{eq:using_equi_nores} is a contradiction, and therefore $\hat{x} > \tilde{x}$.~$\square$

\end{document}